\documentclass[preprint]{aastex}
\shorttitle{SDSS DR6 Galaxy Clusters from AMF finder}
\shortauthors{Szabo et al.}

\begin{document}

\title{An Optical Catalog of Galaxy Clusters Obtained
from an Adaptive Matched Filter Finder Applied to SDSS DR6}
\author{T. Szabo\altaffilmark{1}, E. Pierpaoli\altaffilmark{1}, F. Dong\altaffilmark{2}, A. Pipino\altaffilmark{1,3,4}, J. Gunn\altaffilmark{2} }
 \altaffiltext{1}{Department of Physics and Astronomy, University of Southern
California, Los Angeles, CA 90089, USA}
 \altaffiltext{2}{Department of Astrophysical Sciences, Princeton University, 
Princeton, NJ 08544, USA}
 \altaffiltext{3}{Dipartimento di Fisica, sez.Astronomia, Universit\`{a} 
di Trieste, via G.B. Tiepolo 11, I-34131, Trieste, Italy}
\altaffiltext{4}{Department of Physics and Astronomy, University of California
at Los Angeles, Los Angeles, CA 90095, USA}
\email{tszabo@usc.edu,pierpaol@usc.edu}

\begin{abstract}
We present a new cluster catalog extracted from the 
Sloan Digital Sky 
Survey Data Release 6 (SDSS DR6) using an
adaptive matched filter (AMF) cluster finder.
We identify 69,173 galaxy
clusters in the redshift range 0.045 $\le z <$ 0.78 in
8420 sq. deg. of the sky.
We provide angular
position, redshift, richness, core and virial radii
estimates for these clusters, as well as an error analysis
for each of these quantities.  
We also provide a catalog of more than 205,000 galaxies representing 
the three brightest galaxies in the $r$ band  
which are possible BCG candidates.  
We show basic properties of the BCG candidates 
and study how their luminosity scales in redshift and 
cluster richness.
We compare our catalog with the maxBCG and GMBCG catalogs, as
well as with that of Wen, Han, and Liu. We match between 30\%
and 50\% of clusters between catalogs over all overlapping redshift
ranges.  We find that the percentage of matches increases with
the richness for all catalogs. 
We cross match the AMF catalog with available X--ray data in the same 
area of the sky and find 
539 matches, 119 of which with temperature measurements.
 We present scaling relations between optical and X--ray properties and  
cluster center comparison.
We find that both $\Lambda_{200}$ and $R_{200}$ correlate well with 
both $L_X$ and $T_X$, with 
no significant difference in trend if we restrict the matches 
to flux--limited X-ray samples.

\end{abstract}


\section{Introduction}

Galaxy clusters are the most massive  gravitationally bound systems  
in the Universe. 
Their study has been pursued for many different 
astrophysical and cosmological reasons:
their mass function allows for the determination of the 
 several cosmological parameters, including the mass
density, $\Omega_{m}$, and the matter power 
spectrum normalization $\sigma_{8}$ \citep{Pierpa01,reip02,
seljak02,Pierpa03,dahle06,pedersen07,rines07,Rozo09} as well as 
the dark energy equation of state \citep{Allen08}
and neutrino masses \citep{Wang05}. Furthermore,  cluster surveys 
allow the determination of their clustering properties 
inferring information on the 
large scale structure \citep{bahcall88, postman92,
carlberg96, bahcall97}.  Due to their
high galaxy density, clusters are excellent
laboratories for studying galaxy evolution \citep{dress80,
butch78, garilli99, goto03a, goto03b}.  
Clusters
can also be used as gravitational lenses, providing a way to 
constrain their masses, as well as to 
study distant galaxies \citep{blain99,smail02,metcalfe03,
santos04}.
More recently, cluster abundances and internal structure have 
been invoked as test for modified gravity
\citep{Rapetti08,Diaferio09}. 

Clusters are observed in several bands: in the optical 
through the overdensity of galaxies 
or their color properties, and in the radio and X--ray 
through the emission of the intra--cluster medium.
While optical observations were the first ones to be  
performed, X--ray surveys of the past twenty years have discovered  
hundreds of clusters up to high redshifts ($z \simeq 1$),
while radio surveys aiming at detecting galaxy clusters 
through their Sunyaev--Zel'dovich (SZ) effect 
are underway.
This plethora of data reinvigorates the interest 
in galaxy clusters as it will provide a better understanding of 
cluster physics and the selection function for each 
detection method, therefore improving the precision 
in deriving cosmological constraints.

In this paper, we focus on optical observations of 
clusters in the Sloan Digital Sky Survey (SDSS), 
presenting the cluster catalog constructed from 
an adaptive match filter (AMF) technique \citep{dong08, 
postman96,kepner99, Kim02, White02}.

SDSS provides luminosities in five bands and  
redshift estimates for millions of galaxies
in more than one-fifth of the sky.  This has allowed 
automated algorithms to compile catalogs of clusters objectively
by identifying overdensities in galaxy distributions.
Three such optical cluster catalogs include the maxBCG catalog 
\citep{koester07} constructed from SDSS DR5,
another catalog from SDSS DR6 constructed 
by \citet[][hereafter, WHL]{WHL09}, and the 
Gaussian Mixture Brightest Cluster Galaxy  
catalog \citep[][hereafter, GMBCG]{GMBCG} which is based on SDSS DR7.
The maxBCG catalog makes use of a ridgeline in $g-r$ magnitudes
to determine photometric redshifts and cluster membership.
The catalog by WHL uses a friends-of-friends
algorithm to associate galaxies above a specific luminosity
cut. 
The GMBCG catalog is a follow-up to the maxBCG catalog that
extends the redshift range to $z = 0.55$ and uses statistics
gathered from the maxBCG catalog to refine the cluster finding
procedure.  Given the diverse nature of clusters, one should not expect 
to detect the same object when different methods are applied to the same data.
Comparisons between methods allow for a better 
understanding of selection biases of each method.

We present here the catalog extracted from the SDSS
DR6  by 
applying the AMF technique tailored to handle SDSS 
data \citep[][hereafter, D08]{dong08} and provide 
a comparison with the maxBCG and GMBCG samples. 
We also compare our catalog with the WHL catalog, which was
also derived from DR6 data.  While our finder also uses
overdensities as a starting point for determining the presence
of a cluster, it does not make {\it a priori} assumptions about
the color or number of bright galaxies. 
This approach potentially allows for the 
detection of clusters which do not have a 
bright red galaxy. Moreover, 
 our finder only relies
on extremizing a likelihood function, which provides a
very general method for characterizing 
cluster properties such as their size, richness and core radius 
during their detection.

Clusters may also be detected in the X--ray band, which traces their 
diffuse gas component.
Both X--ray and optical observations are used to determine 
cluster masses and derive cosmology. However, these two bands 
trace different physical components of the cluster, so they represent
quite different probes of the state of the cluster.
 Selection effects for optical and X--ray are different, and 
each method for estimating masses 
may have intrinsic biases and limitations.
Now that big optical cluster samples are available, it is possible to assess 
to what extent optical and X--ray properties are related 
and probe the reliability of 
various methods for mass estimate and cosmology derivation.
In addition, galaxy clusters have now been detected
via their SZ effect \citep{Staniszewski09}.  
Comparisons between the optical properties 
and the SZ effect \citep{Pierpaoli10} using the
Planck Early Release Compact Source Catalog \citep{Planck_ESZ}
will be the subject of a future paper.

As a fist step in this direction, various authors have 
measured scaling relations between 
X-ray observables such as luminosity (L$_{X}$)
and temperature (T$_{X}$)  and optical 
richness \citep{lopes06, rykoff08, popesso05,koester07}.
 While the richness measure used didn't 
seem to correlate well with X--ray luminosity, the 
lensing-derived mass for richness bin showed a 
remarkable correlation with L$_{X}$.
These results were based on shallower  
and smaller optical cluster samples than the one 
presented here as they only relied on a sample on the order of 100 matches. 
Here we provide the X--ray matches for our AMF cluster 
sample with all available 
X--ray cluster data from the BAX database and with the flux limited 
NORAS samples \citep{bohringer00,bohringer04}.
We also measure scaling relations between optical 
richness and X--ray properties for matching clusters.
We find good correlation between optical richness 
and both X--ray luminosity and temperature.

A feature of galaxy clusters that is often considered for its
own study is the brightest cluster galaxy (BCG).
 As part of this work, we
determined the three brightest galaxies in any cluster in
the $r$-band and
include a list of them as potential BCGs.  These can be used
as a further basis for comparison between our clusters and
those in the maxBCG and other catalogs.  
Moreover, BCGs can be exploited as tracers of large scale structure 
in cross correlation analysis with other catalogs 
\citep{Ho09}.
A detailed study of a sub-sample of BCGs
selected to be the brightest and to have z$\le$0.3 
in terms of colors 
and X--ray and UV counterparts is presented in a 
companion paper \citep{Pipino10}.

The organization of this paper is as follows.  In section \ref{data_sect},
we discuss the specifics of the SDSS data used.  We present
the AMF cluster finder and details on its 
application to the SDSS data in section \ref{finder_sect}.
Section \ref{cat_sect} discusses the 
characterization of the galaxy cluster catalog and 
of the related BCGs sample.
In section  \ref{comp_sect} we compare our catalog with the 
maxBCG, WHL, and GMBCG ones, as well as with X--ray  cluster catalogs.
Section \ref{conc_sect} is dedicated to 
the conclusions.

Unless stated otherwise, we assume a $\Lambda$CDM cosmology,
with $\Omega_{m}$~=~0.3 and $\Omega_{\Lambda}$~=~0.7, and
H$_{0}$~=~100 h km~s$^{-1}$~Mpc$^{-1}$.

\begin{figure}
\plotone{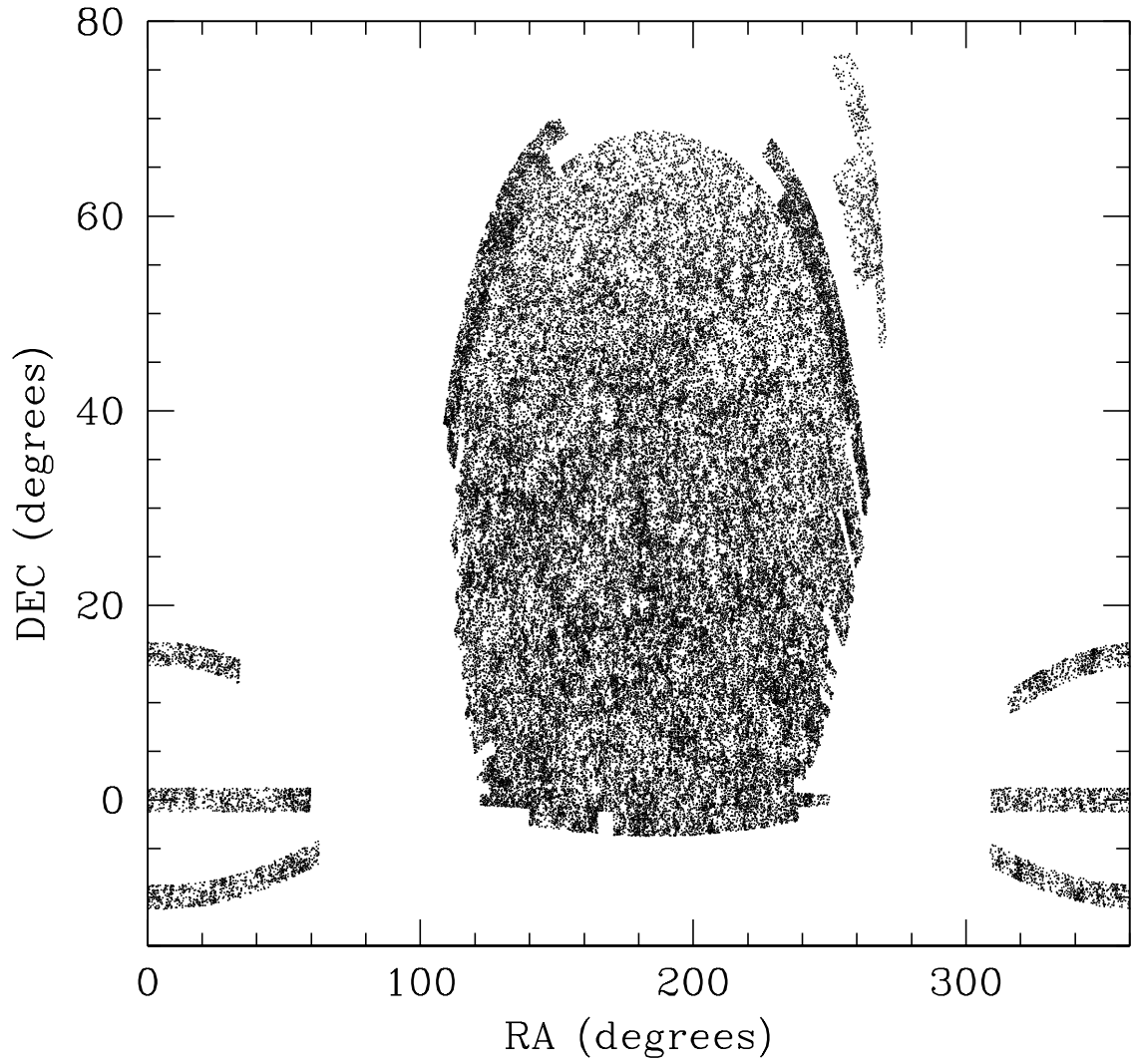}
\caption{Angular positions of the clusters found in SDSS DR6.
\label{sky}}
\end{figure}

\section{SDSS Data \label{data_sect}}

The Sloan Digital Sky Survey \citep{york00,DR6paper} is a five-band CCD 
imaging survey of $10^4$  $\rm{deg}^2$ in 
the high latitude North Galactic Cap and a 
smaller deeper region in the South, followed by 
an extensive multi-fiber spectroscopic survey. The 
imaging survey is carried out in drift- 
scan mode in five SDSS filters (u, g, r, i, z) to a 
limiting magnitude of $r < 22.6$ for 50$\%$ 
completeness \citep{DR5paper}. 
The spectroscopic survey targets 
$8 \times 10^5$ 
galaxies to approximately $r < 17.7$, with a median 
redshift of $z = 0.1$, and a smaller deeper sample of $10 ^5$ 
Luminous Red Galaxies  (LRGs) out to about 
$z =0.5$. 
In this paper we construct a cluster catalog for  
Data Release 6 (DR6)\footnote{\url{http://cas.sdss.org/dr6}} and
chose to use the photometric redshifts provided by  
\citet{oyaizu08}.
In total, we have redshift information for 
$> 98\%$ of 67.6 million galaxies that
covered $\sim$8,420 deg$^{2}$ in stripes 9-39, 42-44,
76, 82 and 86.
In this sample, there are 7.4 $\times$ 10$^{5}$ with 
spectra.  The median error for photometric 
redshift (z$_{photo}$) measurements
is 0.12.
All galaxy measurements were extracted from
the {\tt Galaxy} view on the CasJobs DR6 database
using the {\tt FLAGS} as described in 
Appendix \ref{phot_flags_app}
to limit our sample to galaxies
with good photometry.
We use the {\tt cc2} photo-z estimates in order to avoid 
a possible bias from applying  
the luminosity function twice, once in the choice of photometric
redshifts and again as part of the AMF likelihood calculation. 

It is important to understand the errors in photometric redshifts
when using our technique, as we explain below. We used the photometric redshifts 
provided by \citet{oyaizu08} in preference to those from \citet{csabai03}, even though the latter were available for the whole
of DR6, because   the errors on z$_{photo}$ for \citet{csabai03} 
were underestimated. The Oyaizu {\em et al}  
errors on z$_{photo}$ are much closer to
the errors $|z_{spec} - z_{photo}|$ estimated from the 
sub-sample of galaxies which have
spectroscopic redshifts. 
The \citet{csabai03} estimates would bias our 
redshift filter in such a way that 
would eliminate galaxies that should be considered as 
possibly belonging to a given cluster.  The width of the
filter is determined by the error estimate in redshift. 
Also, the \citet{oyaizu08} sample is truncated at an
$r$-band magnitude of 22.  This is important, as the behavior
of the luminosity function that we use is not well-understood
above this magnitude.  In addition, the coverage for the 
region of the sky over which we are constructing our catalog
is much more uniform for the redshifts provided by \citet{oyaizu08}
than for those provided by \citet{csabai03} for SDSS DR6. Simulations using
a mock galaxy catalog indicated that this procedure should lead to a catalog
that is $>$90$\%$ complete at least to $z=0.4$ (D08).

The current catalog is constructed using $r$-band data information only,
although the code could be extended to make use of 
information in all bands.
Figure \ref{sky} shows the distribution of clusters in
our catalog on the sky.  
We will assess data based on the continuity and completeness
of its redshift estimates for future releases of 
AMF cluster catalogs.

\section{The Cluster Finder \label{finder_sect}}
\subsection{Method}
The cluster finder that we employ is based on 
the matched filter concept \citep{postman96,kawasaki98,kepner99,Kim02,White02}.
A complete description of the method used here as well as 
results of testing on mock data
is available in D08.

The matched filter technique adopted here is a likelihood method 
which identifies clusters by convolving the optical galaxy 
survey with a set of filters based on a modeling of 
the cluster and field galaxy distributions. A cluster radial 
surface density profile, a galaxy luminosity function, and 
redshift information (either photometric or spectroscopic 
when available) are used to construct filters in 
position, magnitude, and redshift space, from which a cluster 
likelihood map is generated.  The peaks in the map thus 
correspond to candidate cluster centers where the matches 
between the survey data and the cluster filters are optimized. 
The algorithm automatically provides the probability for 
the detection, best-fit estimates of cluster properties 
including redshift, radius and richness, as well as 
a framework for membership 
assessment for each galaxy.

The cluster finding algorithm computes the likelihood on a grid 
whose points are centered 
on the positions of galaxies.  The finder uses an iterative
procedure that determines which galaxy maximizes the
difference in likelihood from the previous step that a cluster
exists that is centered on that galaxy.  This cluster is then
added to the list of identified clusters, and the procedure
continues with the remaining galaxies.
The parameters of the new cluster 
(redshift $z_c$, core radius $r_c$ and 
richness $\Lambda_{200}$) are varied in order 
to maximize its likelihood increment. 
The radius $R_{200}$ of the cluster 
is determined by computing at which distance from the central 
galaxy the overdensity 
of galaxies is 200 times the critical density, assuming the 
average galaxy density is representative of 
the mean mass density.
The richness is the total luminosity within
$R_{200}$ in terms of $L^{*}$, 
where $L^{*}$ evolves
passively as a function of redshift, brightening by
0.8 mag from $z = 0$ to $z = 0.5$ \citep{Loveday92, Lilly95b, Nagamine01,
 Blanton03, Loveday04, Baldry05, Ilbert05}.  We assume that
$L^{*}$ does not vary with the cluster richness, but see \citet{hansen09} and section 4.3.
The cluster finding procedure is stopped when 
the natural logarithm of the likelihood  
for any new cluster is below zero.  (For the remainder of
this paper, any mention of a value of the likelihood
refers to the $\ln(likelihood)$.)
From this sample of cluster candidates, all clusters with
richness $\ge$ 10 are chosen for recentering, which 
is described below.

Since the matched filter technique does not explicitly use the 
information about the red sequence to select clusters as  
in some color-based cluster-finding methods \citep{Annis02,
Miller05,koester07}, it can theoretically detect clusters 
of any type in color, and is not restricted only to old, red 
E/S0 galaxies. Such clusters likely dominate the cluster 
population, but may not constitute all of it especially as 
one probes systems of
lower richness and at higher redshifts.
We refer the reader to \citet{Pipino10} for further discussion on this issue.

\subsection{Completeness and Purity Estimate}

By applying this cluster finder to simulations, D08  
demonstrated that, with a richness cut at $\Lambda_{200} \ge$ 20,
the results of this finder are over 95$\%$ complete for objects
with $M_{200} >$ 2.0 $\times$ 10$^{14}h^{-1}M_{\sun}$
and $\sim$85$\%$ complete for objects with
$M_{200} >$ 1.0 $\times$ 10$^{14}h^{-1}M_{\sun}$ in the
redshift range 0.10 $<$ z $<$ 0.45.  These authors also showed
 that the finder produces a catalog that is over 95$\%$
pure for clusters with $\Lambda_{200} >$ 30 and 90$\%$ pure
for clusters with $\Lambda_{200} >$ 20 over all redshifts
z $<$ 0.45.  
No testing has been performed on deeper simulations, 
so that we are not in the position 
of making statements of purity and completeness 
for any redshift beyond $z=0.45$.

\subsection{Recentering of Clusters \label{recenter}}

The initial run of the finder produces clusters that are
centered on observed galaxies.  
However, there is no reason to assume that the 
actual center of mass for the cluster 
should necessarily be located on a  galaxy.

In general, finding the correct  
center of mass of a cluster is important for several reasons:
$i)$ for mass estimation with the lensing effect, 
as clusters are stacked one on top of the other;
$ii)$ for comparison with measurement of the cluster in other bands,
$iii)$ for assessing velocity dispersions and therefore 
kinetic mass estimates of the cluster; 
$iv)$ for the purpose of cross--correlations with other surveys and CMB maps. 
Moreover, trying to determine cluster's properties 
that rely on a spherical model while not using 
the correct center may lead to determining them poorly and 
assigning a bigger error to them.

For all these reasons, after the clusters have been identified, 
it is necessary to relax the hypothesis
that clusters are centered on a galaxy and  
refine the center position as well as the 
characterizations of the clusters.
We therefore recompute the cluster's 
likelihood for each cluster in the sample by varying 
the new hypothetical center on a fine grid of  
resolution 1 h$^{-1}$ kpc and whose extent
is the core radius, r$_c$.
 About 37$\%$ of the clusters have their center
shifted by at least 1 h${-1}$ kpc; 
the average displacement for clusters whose center
is measurably displaced is 65$^{+32}_{-58}$ h$^{-1}$ kpc.  All other 
quantities ($\Lambda_{200}$, $R_{200}$, $z$, and $r_{c}$)
are recomputed for the new angular position.  

\subsection{Blended clusters}

The procedure outlined above finds 69,173 clusters with
$\Lambda_{200} \ge$ 20 in DR6.  These clusters are distributed
over 66,231 unique sites in angular and redshift space. 
To determine unique sites, we first determine which galaxies
have a high likelihood of belonging to each cluster.
We then determine the three brightest galaxies in the
$r$-band for each cluster.  Clusters which share any of
the three brightest members are said to belong to the
same site.  The cluster with the highest likelihood at
a given site is said to be the primary cluster.  Since
the richness scales as the likelihood, this is also the
richest cluster at a site.

To assign galaxies to a particular cluster, 
we make a list of all galaxies ($i$) that may
belong to a cluster ($k$) such that the angular displacement
of the galaxy from a cluster center is less than 
$R_{200,k}/d_{A}(z_{i},\Omega_{m},\Omega_{\Lambda})$
and $|z_{k} - z_{i}|$ $<$ 3$\sigma_{z}(i)$.  
Here, $\sigma_{z}(i)$ is the estimate of the error in
photometric redshift assigned to the galaxy in SDSS DR6 
and d$_{A}$ is the angular diameter distance to the galaxy, $i$.
This is not a
final membership list for the set of clusters, but a set
of galaxies that could contribute, however minimally, to 
the likelihood and richness of a cluster.  
For galaxies within $R_{200}$ of the center of the cluster,
we then find the likelihood of the cluster when it is
missing galaxy $i$.  The difference between the likelihood
of the cluster with all members included and the likelihood
when missing galaxy $i$ is referred to as the
likelihood difference, $\mathcal{L}_{i}$.
We then assign memberships of galaxies 
to a cluster by selecting 
galaxies that lie within $R_{200}$ from the cluster's center,
and have at least $\mathcal{L}_{i}=1$.  With this threshold,
we can assign member galaxies to $>$99.4$\%$ of the clusters.
This cutoff preserves galaxies with a spectroscopic redshift
measurement that is close to the redshift of the cluster
as cluster members.
Selecting a cutoff of $\mathcal{L}_{i}>1$ quickly reduces the
percentage of clusters with members assigned, and a cutoff
$<1$ includes too many bright galaxies with no redshift
estimate.  This choice of cutoff also reduces the 
median value of the error in photometric redshift for galaxies
chosen to be cluster members to 0.062.
 We do not apply a prescribed redshift cut because 
 the high likelihood cut applied already ensures that the selected galaxies 
 are close to the cluster's center.
The galaxies selected with this criterion are 
included in the catalog (see Table \ref{app_mem}).

By using this method to determine the brightest galaxies in a cluster
and requiring that the brightest galaxies can only belong to one site,
we determine that $<$5$\%$ of sites have more than one cluster attributed
to them.
We choose to retain these blended clusters in the catalog as possible
subjects of cluster mergers.
We also want to know how these sites compare with detections in
other bands.  Lastly, these blended clusters provide insight 
into the AMF filter detection method.

\section{The AMF Catalog \label{cat_sect}}

In this section we present the AMF cluster catalog
and the associated BCG. We present its main properties 
including richness, redshift distribution, core radius 
with associated errors.
We describe how the BCG catalog is assembled and 
present its main properties in the $r$ band.
A description on how  the data are released and how to
retrieve them is in the Appendix.

\subsection{Main Properties}

The catalog contains 69,173 clusters with
$\Lambda_{200} \ge$ 20 over an area of $\sim$8,420 deg$^{2}$. 
Results of the cluster finder applied to simulations (D08) 
associate this richness 
threshold with an approximate mass of $4\times 10^{13} M_\odot$, 
according to:

\begin{equation}
\Lambda_{200} 
= (47.2\pm4.1)\times \bigl(\frac{M_{200}}{10^{14} h^{-1} M_{\sun}}\bigr)^{1.03\pm0.04}
\label{eq:scr}
\end{equation}.

Using this result from the simulation as a guideline,
we detect that
there are about 5000 clusters in our new AMF SDSS
catalog with a mass above 
$10^{14}h^{-1}M_\odot$ (roughly corresponding to 
richness 50); 2700 of which are nearer than $z=0.4$.
It should be kept in mind, however, that eq.\ref{eq:scr} was 
derived on the basis of simulations
where the galaxy colors were assigned  
to dark matter particles, so this result  may not necessarily
correspond precisely to what we observe in nature. 
The exponent in the relation is different from one only by statistical error,
and merely reflects the mass--to--light ratio of cluster galaxies assumed in the simulations.
A proper calibration of the mass--richness relation 
should be made by evaluating masses 
for this sample possibly  with different methods. 
This issue will be properly addressed in 
a future paper.

The position of the clusters in the sky is represented in fig.~\ref{sky}.
It is clearly not homogeneous, as photometric 
redshifts are not determined in a uniform way on the sky 
The density of clusters varies from stripe to stripe,
with an average number of clusters per deg$^{2}$ of
7.2. 
The distribution of  cluster density per deg$^{2}$
varies between 5.4--6.0 (stripes 43-44,76,82, and 86), 
6.0--7.2 (stripes 9-37) and $>$ 9 (stripes 38-39).
The smallest density belongs to stripe 42 which has
4.04 clusters per deg$^{2}$, and the largest density
is in stripe 38 at 9.37 clusters per deg$^{2}$.
We are confident that the variations in cluster
density in stripes 9--37 are due to large-scale
structure.  More work is needed to understand the
density variations in other stripes.

\begin{figure}
\plotone{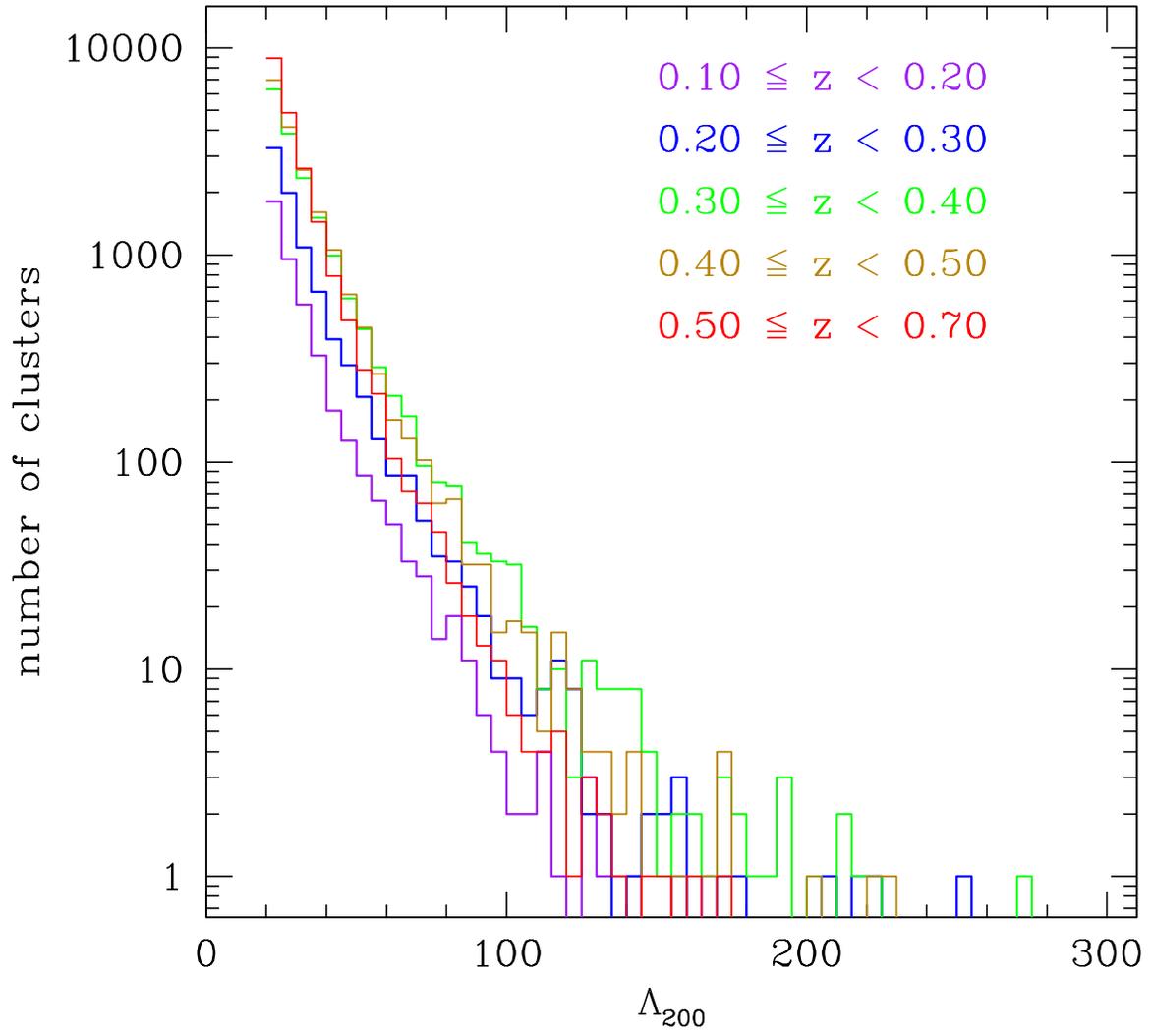}
\caption{Distribution of AMF clusters in richness per redshift bin.
\label{amf_rh_hist}}
\end{figure}

The distribution of our clusters in richness for different 
redshift bins is shown in 
fig.~\ref{amf_rh_hist}. As expected,  the ratio of 
high richness objects over low richness ones decreases 
for increasing redshift. The richest clusters have richness values 
168, 254, 270, 226, and 172 in subsequent 
redshift bins  with $\Delta z = 0.1$ starting at $z=0.1$.
Note that the catalog is clearly not complete for z$>$0.5.  For
this reason, the considerable drop in the number of very rich clusters
compared to the low richness ones, although expected to a certain extent 
for structure formation reasons, may in fact be due to 
limitations of our searching method.
A proper assessment of this issue would require 
comparison with other cluster finders 
performances at high redshifts or better understanding 
of this cluster finder on SDSS deeper data.

\begin{figure}
\plotone{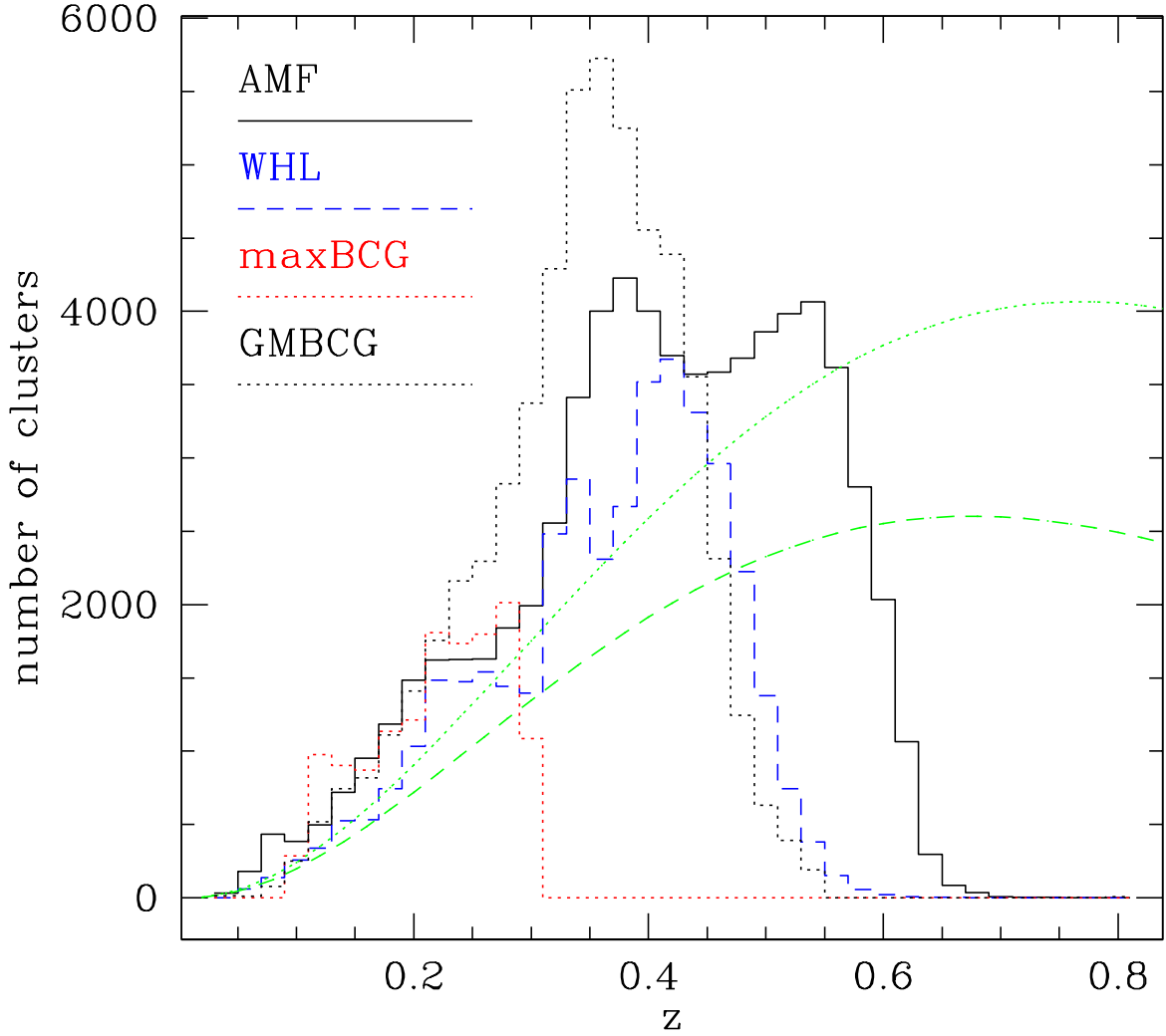}
\caption{
Redshift Distribution of AMF (solid line),
WHL (dashed line, blue),
maxBCG (dotted line, red), and 
GMBCG (dotted line, black) clusters.  The green lines represent the expected
number of clusters for $\sigma_{8} = 0.8$ (dashed) and 
$0.9$ (dotted) and a
mass threshold of $4 \times 10^{13} h^{-1} M_{\sun}$. 
The value of h = 0.70.  Note that the richness
cutoff for the maxBCG and GMBCG catalogs is lower than that
for the AMF catalog. \label{amf_z_hist}}
\end{figure}

The redshift distribution of the cluster catalog  
is presented in fig.~\ref{amf_z_hist},  together 
with the distribution of the maxBCG catalog.  The green, 
dotted lines in this
plot represents the number of clusters expected in each bin
for $\sigma_{8} =$ 0.8 and 0.9, and a mass 
threshold of $M = 4 \times 10^{13} h^{-1} M_{\sun}$,
which corresponds to $\Lambda_{200} \sim $20 according 
to eq. \ref{eq:scr}.
The cluster sample  redshift distribution has a peak at 
$z=0.38$ and then roughly flattens until $z=0.55$. 
However, clusters are found  out to redshift 0.78.
The total number of clusters with redshift below $0.3, 0.4, 0.5, 0.78$ is, 
respectively: 13593, 30814, 49198, and 69173.  

\begin{figure}
\plotone{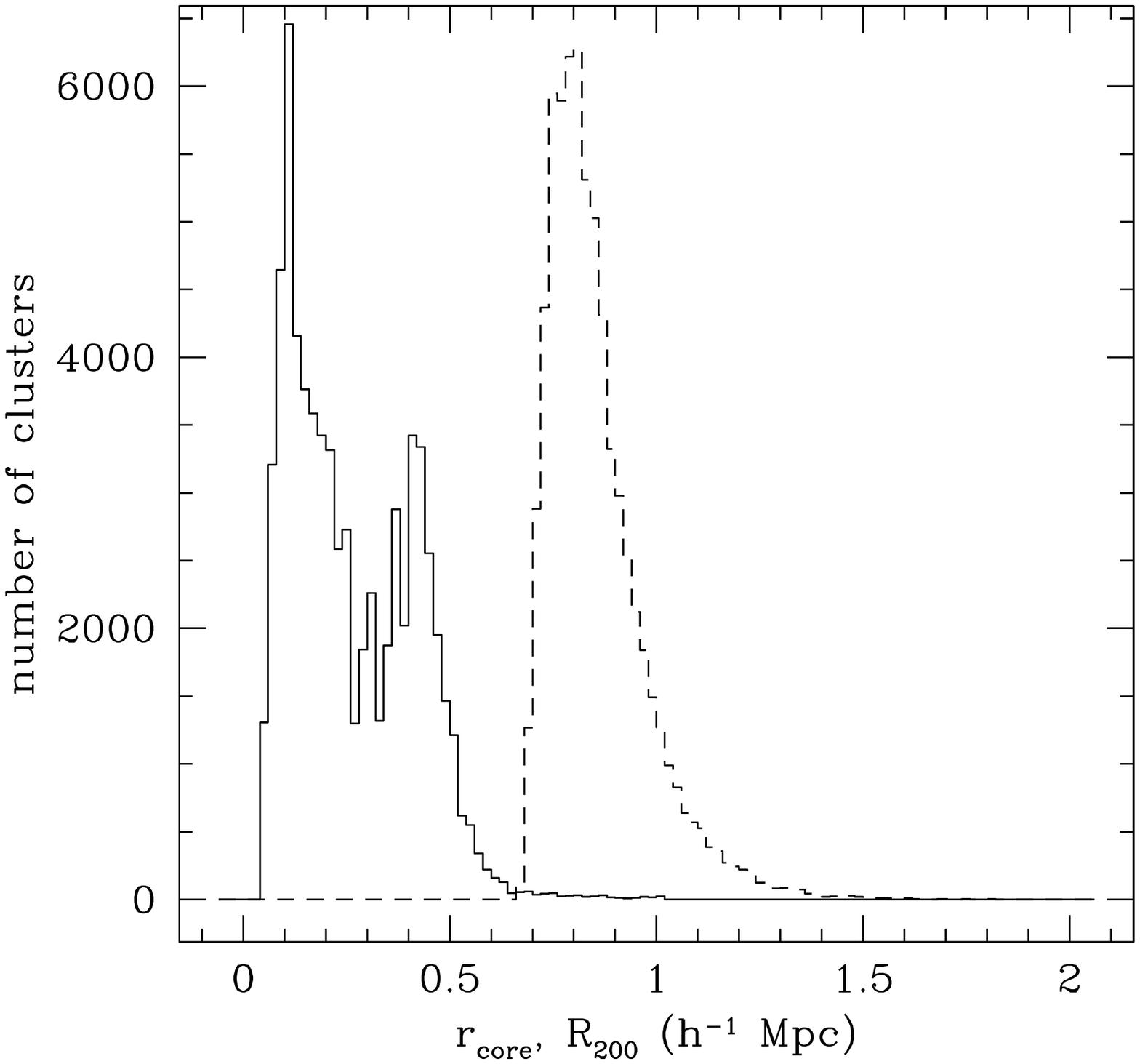}
\caption{Distribution of AMF clusters as a function
of core radius (solid line) and R$_{200}$ (dashed line). 
\label{amf_rc_hist}}
\end{figure}

Histograms of the core radius and R$_{200}$ are presented in fig.~\ref{amf_rc_hist}.  
The distribution in radii is fairly peaked around 0.8 h$^{-1}$ Mpc, 
while the distribution of  core radii  is broader with some suggestion of bimodality.
Most clusters have a
core radius between 80 and 240 h$^{-1}$ kpc.

\begin{figure}
\plotone{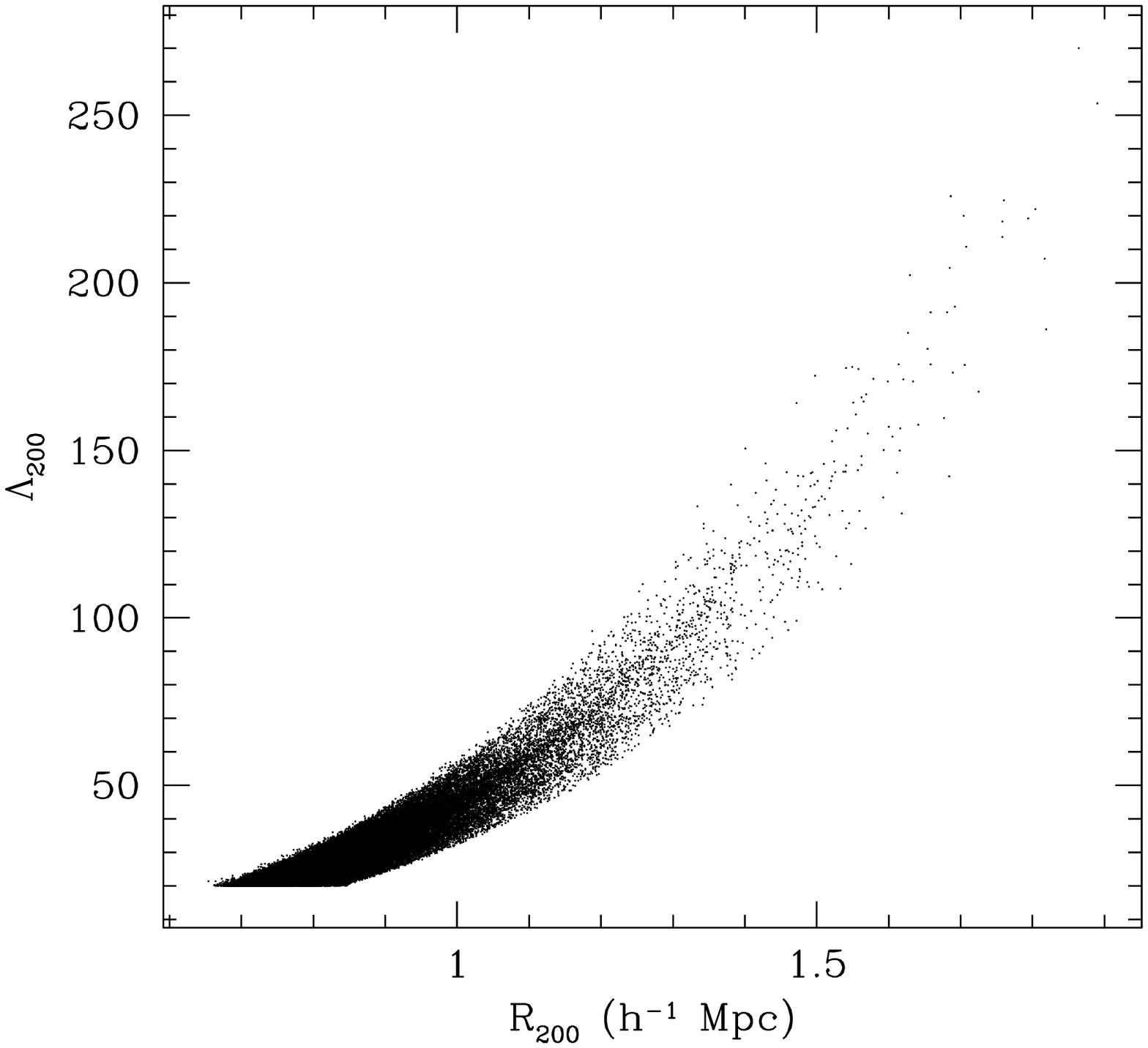}
\caption{Plot of $\Lambda_{200}$ vs. R$_{200}$ for all
69,173 clusters. \label{amf_rh_r200}}
\end{figure}

Finally, we notice that the estimated radii correlate very 
well with richness (see fig. \ref{amf_rh_r200}), as expected 
from tests on simulations (D08). The relationship between 
$\Lambda_{200}$ and R$_{200}^{3}$ is linear, with a dispersion
relation given by:
\begin{equation}
\sigma_{R_{200}} = (0.019\pm0.005)R_{200} + (0.026\pm0.006),
\end{equation}
where $\sigma_{R_{200}}$ is measured in h$^{-1}$ Mpc.

\subsection{Error Determination \label{err_det_sect}}
For each cluster found, the catalog reports the maximum 
likelihood value for the
three main parameters ($\Lambda_{200}$, $z$, and $r_{c}$), as well as for  
the determination 
of $R_{200}$.
The likelihood determination procedure also allows us 
to find the errors 
of the varied quantities
($\Lambda_{200}$, $z$, and $r_{c}$)
as well as for the angular positions of clusters.
In order to determine errors, the likelihood for each cluster 
has been recomputed on a finer 
grid and only considering galaxies within 
2$R_{200}$ and 3$\sigma_{z}$ of the cluster
center.  This larger distribution in angular space
allows us to vary the location of the center to 
determine confidence regions in angular space.
Errors on each quantity were found by
exploring the likelihood surface in two of the parameters
(e.g., $r_{c}$ and $\Lambda_{200}$) while keeping the
others fixed.  The boundaries of the 68$\%$, 90$\%$, 95$\%$,
etc. confidence regions 
are determined by looking for the extrema of the parameters
which give a difference in likelihood from the maximum 
value.

Errors for each cluster are reported in the catalog,
and include the extrema of the 68$\%$ and 95$\%$ confidence
range for each quantity. 
Table \ref{rh_table} gives errors as a function
of $\Lambda_{200}$, table \ref{z_table} as a
function of redshift, and table \ref{rc_table} as
a function of core radius. In these tables, the
interquartile range is provided for each error
estimate.

Errors in $\Lambda_{200}$ were measured as a function
of $\Lambda_{200}$, $z$, and $r_{c}$.  The error in richness
increases as richness decreases.  This is to be expected,
as the addition or subtraction of one galaxy creates a 
larger percentage difference in smaller clusters.  Typical
values range from around 45$\%$ for clusters with 
$20 \le  \Lambda_{200} < 30$ to 15 to 20$\%$ for clusters
with $\Lambda_{200} >$ 150.  Errors in $\Lambda_{200}$ change
very little as a function of $z$, though there is a small
decrease with increasing redshift.  Typical errors are
$\sim$45$\%$, as the smallest clusters are the ones that
are most prevalent.  The error in $\Lambda_{200}$ is 
anti-correlated with the size of the core radius estimate.
Clusters with core radii $<$0.1 h$^{-1}$ Mpc have errors in
$\Lambda_{200}$ around 50$\%$, and those with core radii
$\ge $~0.5 h$^{-1}$ Mpc have richness errors typically
in the vicinity of 37$\%$.  We also find that the average
core radius grows steadily from 0.23 h$^{-1}$ Mpc
for clusters with 20 $\le  \Lambda_{200} < $30 to 
0.43 h$^{-1}$ Mpc for clusters with $\Lambda_{200} \ge $ 100,
so this anti-correlation is expected.

Several studies have indicated that the fit of an NFW profile
for optical clusters tends to have a large error in core radius, 
$r_{c}$, and that in some cases, a function with
either a core or a cusp in the density at the 
center of the cluster gives an adequate 
description \citep{katgert04,biviano06}.
Our error analysis similarly finds the core radius to
be poorly constrained in most cases.  As with 
the other parameters, the error in core radius is smallest
for the richest clusters, with errors falling between
14$\%$ and 36$\%$ for half of the 327 clusters with
$\Lambda_{200}  \ge$ 100.  For the 43,600 clusters with 
20 $\le  \Lambda_{200} <$ 30, the endpoints of this
range shift to 46$\%$ and 87$\%$.  The redshift of the
cluster also has a strong effect on the typical error
value for the core radius.  Nearby clusters ($z \le  0.2$)
have errors that are about the same size as the 
estimate for the value of the core radius, while the
errors drop to about half the core radius (40--70$\%$)
for clusters with $z \ge  0.4$.  There is no clear
trend in the error in core radius as a function of core
radius, except that the values are greater than 40$\%$
for all clusters with r$_{c} <$~0.5~Mpc~h$^{-1}$.

As we discussed in section \ref{recenter}, our clusters are not
required to be centered on a galaxy.  This allows us to explore
a generalized position in angular space to determine how
accurately we found the cluster center.  The errors are
determined by varying the position of the center over the
core radius of the cluster; therefore, it is important to 
note any dependence of error on r$_{c}$ first.  For
r$_{c} \ge $ 0.2 h$^{-1}$ Mpc, the values of the boundaries
for the middle 50\% of the error values are consistently
about 18$\%$ of R$_{200}$ for the low end and 
about 32$\%$ for the high end.
This is probably the best error estimate for clusters with
r$_{c} <$ 0.2 h$^{-1}$ Mpc as well, as it appears that
varying the position over the core radius limits the maximum
extent to which you can sample the likelihood surface for 
this set of clusters.  
Errors in angular
position anti-correlate with the richnesses of the clusters.
The richest clusters will have errors in position typically 
$<$10$\%$ of R$_{200}$,
while the poorest clusters in our catalog have position errors
usually in the range of 20--25$\%$ of R$_{200}$.  As a function 
of redshift, the smallest typical value for angular error is
consistently 15$\%$ of R$_{200}$, except for $z < 0.1$, where it
is 12$\%$.  The largest typical error in angular position 
increases from 23$\%$ of R$_{200}$ for $z < $0.1 to 32$\%$ for
$z \ge $0.5.

\begin{figure}
\plotone{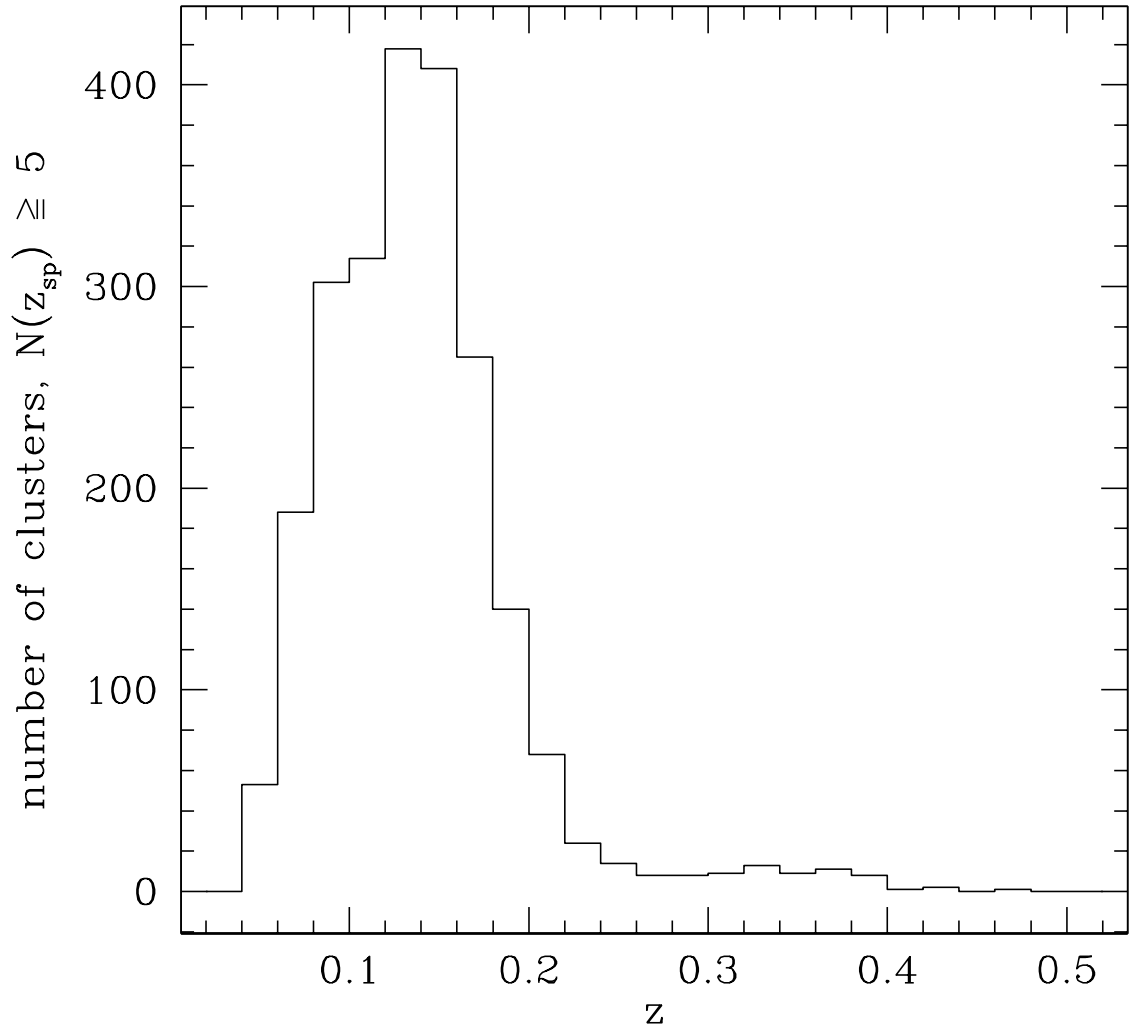}
\caption{Distribution in redshift of clusters that have at
least 5 galaxies with spectroscopic redshifts.\label{z_zsno}}
\end{figure}

\begin{figure}
\plotone{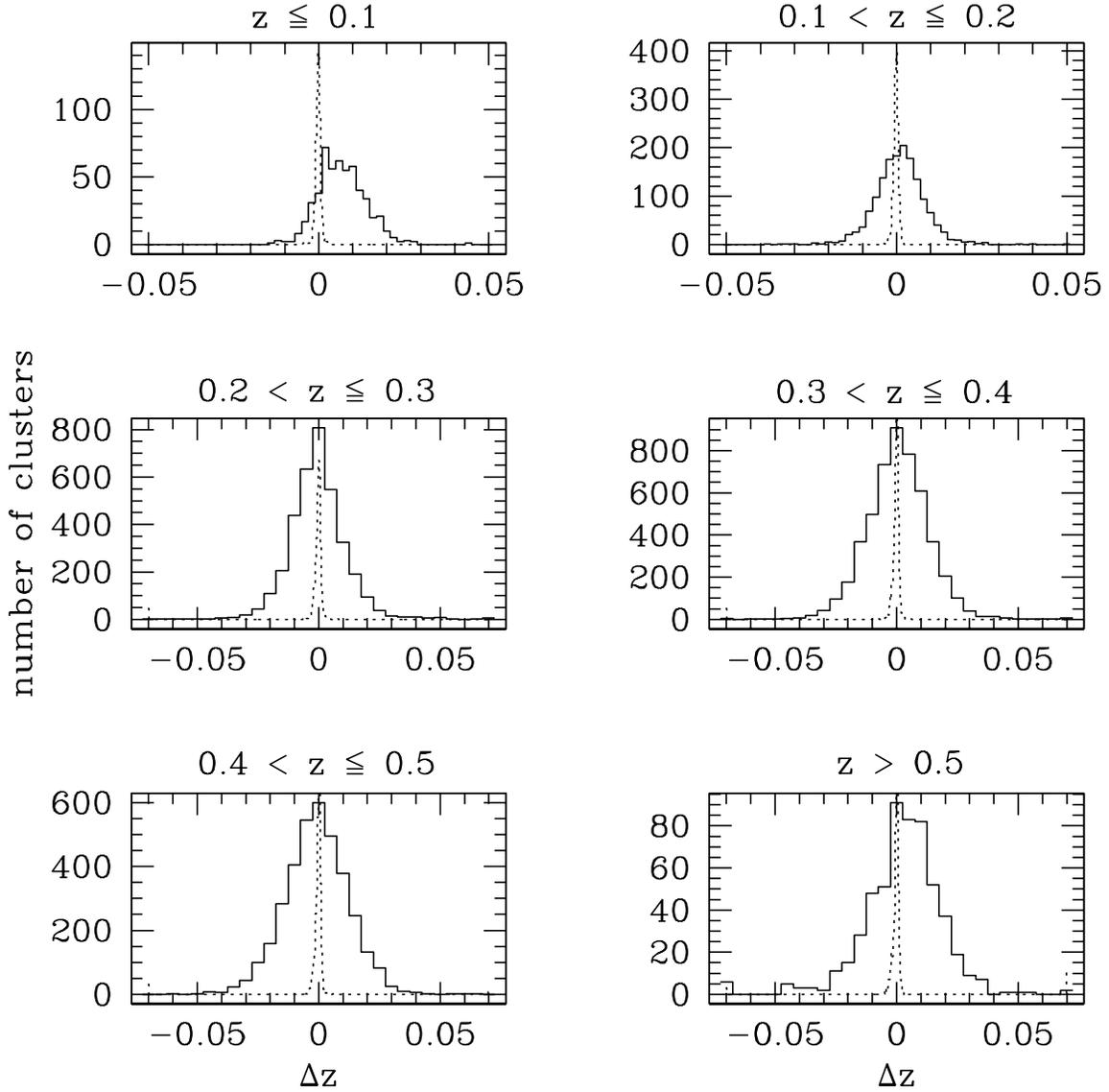}
\caption{Difference between the clusters' redshift 
realizing the  maximum  likelihood ($z_{cl}$)
with one obtained from only photometric redshifts ($z_p$, solid line) and one
from only spectroscopic redshifts ($z_s$, dotted line).
\label{zpzs_comp}}
\end{figure}

As for redshift estimates, the cluster finder assigns a redshift,  
$z_{cl}$, according to the
kind of redshift measurements available for the galaxies in the cluster.
The majority of distant clusters have no galaxies with spectroscopic 
redshift measurements, so that $z_{cl}$
gets estimated only from photometric measurements.
Here we would like to assess the accuracy of this procedure.
To this aim, 
we analyze errors in estimated cluster redshift, $z_{cl}$,
as obtained 
from the maximum likelihood procedure
by comparing photometric and spectroscopic 
redshift determinations for 
the union of two sets of clusters: 
i) all clusters 
that have at least five galaxies with spectroscopic measurements
(mainly located at $z \le 0.2$, see fig.~\ref{z_zsno}),
and ii)  clusters at $z >$ 0.2 with  at least one 
of their three brightest members in the $r$-band  having a spectroscopic measurement and $\mathcal{L}_i \ge $ 1 (see also WHL).
For these clusters  we
compute the  redshift  $z_p$ by only using the photometric redshifts  of all galaxies.
We also compute the redshift estimate $z_s$ from the spectroscopic
redshifts, which we make correspond  to the BCG redshift in the latter case.
These redshift values are then compared to the one obtained
using the maximum likelihood method used in the AMF finder ($z_{cl}$).
Results are reported in Table  \ref{zpzs_table} and Figure  \ref{zpzs_comp}.

We compute 
$\Delta_z \equiv z_{cl} - z_p$,
that is the  difference between redshift of the maximum
likelihood point (which make use of the 
spectroscopic redshifts when available)
with that obtained from the product of
Gaussians using only the photometric redshifts, 
 as a function of redshift bin.
We also determine $z_{cl} - z_s$ for each cluster in the sample 
(see Table \ref{zpzs_table}).
For the redshift range where there are many clusters
with five spectroscopic redshifts ($z \le$ 0.2),
the width of the distribution of differences,
$\sigma_{z_{cl} - z_p}$, is about 0.007.
For
$z \ge 0.2$
 we find the width of the
distribution to be 0.013, except for the most
distant bin ($z >$ 0.5), where the width is 0.022.
Except for the lowest redshift range ($z <$ 0.1), the 
distribution of differences has an average $|\Delta_z | <$ 0.002
and is considered to be unbiased
(see Figure \ref{zpzs_comp}).
For clusters with z $<$ 0.1, there is a distinct bias between
redshifts determined spectroscopically and those determined
photometrically which makes clusters with only photometric
redshifts appear $\Delta z \simeq 0.007$ more distant on average.
However, of the 838 clusters with z $<$ 0.1, 64$\%$ have
five or more galaxies with spectroscopic redshift measurements
associated with them.

We use the width of the distributions,
$\sigma \equiv \sigma_{z_{cl} - z_p}$, as a characteristic
error for that redshift range 
and 
compare
$|\Delta_z|$ with 3$\sigma$.  Over the range
$z \ge  0.1$, the percentage of clusters 
outside 3$\sigma$ is less than 1.5\% per redshift bin.
This small percentage of clusters with larger
redshift differences is considered to be
characteristic of the entire sample, as
the distribution of {\tt cc2} photometric
redshift estimates with $r >$ 20 from \citet{oyaizu08}
mirrors the distribution of spectroscopic
redshifts from their sample.
 No clusters  in the $z \le  0.2$ range have
$|z_{cl} - z_s| > \sigma$.  From analyses focusing
on redshift error versus number of spectroscopic measurements,
we know clusters with at least five spectroscopic
measurements will have an average error in 
redshift that is about equal to the
velocity dispersion of the constituent galaxies, or
$|\Delta z| < 0.002$ with no bias.
In all cases, 
the $z_{cl} - z_s$ is sharply peaked, even for
clusters with only one spectroscopic measurement (see dotted lines
in Figure \ref{zpzs_comp}).  Projection effects place
a BCG candidate outside of $\sigma$ in $<$ 1.7\%
of the clusters from $z =$ 0.2 to $z =$ 0.5.  For
$z >$ 0.5, this percentage increases to 2.9\%. 
This analysis allows us to state that $\sigma_{z_{cl} - z_p}$
is a reasonable value to use for the error of 
the redshift of clusters in each bin.

\begin{deluxetable}{llcccccc}
\tabletypesize{\footnotesize}
\tablecolumns{8}
\tablewidth{370.0pt}
\tablecaption{Errors as a function of richness\label{rh_table}}
\tablehead{
\multicolumn{1}{c}{} & \multicolumn{1}{c}{} &
\multicolumn{2}{c}{$\Lambda_{200}$} & \multicolumn{2}{c}{r$_{c}$} &
\multicolumn{2}{c}{position ($\%$ of R$_{200}$)} \\
\multicolumn{1}{c}{$\Lambda_{200,min}$} &
\multicolumn{1}{c}{$\Lambda_{200,max}$} &
\multicolumn{1}{c}{$Q_1$} & \multicolumn{1}{c}{$Q_3$} &
\multicolumn{1}{c}{$Q_1$} & \multicolumn{1}{c}{$Q_3$} &
\multicolumn{1}{c}{$Q_1$} & \multicolumn{1}{c}{$Q_3$}
}
\startdata
100 & \nodata & 20.2$\%$ & 23.9$\%$ & 14.1$\%$ & 36.0$\%$ & 6.5$\%$ & 12.7$\%$ \\
80 & 100 & 25.4$\%$ & 27.6$\%$ & 19.5$\%$ & 43.2$\%$ & 8.2$\%$ & 15.9$\%$ \\
60 & 80 & 28.4$\%$ & 32.1$\%$ & 23.0$\%$ & 51.5$\%$ & 9.9$\%$ & 18.0$\%$ \\
50 & 60 & 32.1$\%$ & 35.1$\%$ & 27.5$\%$ & 61.6$\%$ & 11.1$\%$ & 20.4$\%$ \\
40 & 50 & 35.8$\%$ & 39.6$\%$ & 32.3$\%$ & 66.8$\%$ & 12.5$\%$ & 22.5$\%$ \\
30 & 40 & 39.6$\%$ & 44.0$\%$ & 39.1$\%$ & 74.2$\%$ & 14.2$\%$ & 26.3$\%$ \\
20 & 30 & 42.5$\%$ & 50.0$\%$ & 46.4$\%$ & 87.3$\%$ & 17.0$\%$ & 31.9$\%$ \\
\enddata
\tablecomments{Ranges are listed for $\Lambda_{200,min} \le  
\Lambda_{200} < \Lambda_{200,max}$.
$Q_1$ and $Q_3$ are the lower and upper quartiles, respectively. 
The minimum value of 
r$_{c}$ in the catalog is 0.059 Mpc, and the maximum value of 
$\Lambda_{200}$ is 270.15.}
\end{deluxetable}

\begin{deluxetable}{llcccccc}
\tabletypesize{\footnotesize}
\tablewidth{300.0pt}
\tablecolumns{8}
\tablecaption{Errors as a function of redshift\label{z_table}}
\tablehead{
\multicolumn{1}{c}{} & \multicolumn{1}{c}{} &
\multicolumn{2}{c}{$\Lambda_{200}$} & \multicolumn{2}{c}{r$_{c}$} &
\multicolumn{2}{c}{position ($\%$ of R$_{200}$)} \\
\multicolumn{1}{c}{z$_{min}$} &
\multicolumn{1}{c}{z$_{max}$} &
\multicolumn{1}{c}{$Q_1$} & \multicolumn{1}{c}{$Q_3$} &
\multicolumn{1}{c}{$Q_1$} & \multicolumn{1}{c}{$Q_3$} &
\multicolumn{1}{c}{$Q_1$} & \multicolumn{1}{c}{$Q_3$}
}
\startdata
\nodata & 0.1 & 41.8$\%$ & 52.3$\%$ & 74.5$\%$ & 149$\%$ & 12.0$\%$ & 23.0$\%$ \\
0.1 & 0.2 & 41.8$\%$ & 51.5$\%$ & 64.3$\%$ & 116$\%$ & 15.6$\%$ & 23.8$\%$ \\
0.2 & 0.3 & 40.3$\%$ & 49.3$\%$ & 49.2$\%$ & 91.9$\%$ & 15.3$\%$ & 26.2$\%$ \\
0.3 & 0.4 & 38.8$\%$ & 47.8$\%$ & 43.0$\%$ & 79.7$\%$ & 15.3$\%$ & 28.6$\%$ \\
0.4 & 0.5 & 38.1$\%$ & 47.0$\%$ & 39.5$\%$ & 74.2$\%$ & 15.2$\%$ & 30.3$\%$ \\
0.5 & \nodata & 38.1$\%$ & 47.0$\%$ & 37.9$\%$ & 69.7$\%$ & 15.3$\%$ & 32.4$\%$ \\
\enddata
\tablecomments{Ranges are from z$_{min} <$ z $\le$ z$_{max}$.
$Q_1$ and $Q_3$ are the lower and upper quartiles, respectively.}
\end{deluxetable}

\begin{deluxetable}{llcccccc}
\tabletypesize{\footnotesize}
\tablewidth{330.0pt}
\tablecolumns{8}
\tablecaption{Errors as a function of core radius\label{rc_table}}
\tablehead{
\multicolumn{1}{c}{} & \multicolumn{1}{c}{} &
\multicolumn{2}{c}{$\Lambda_{200}$} & \multicolumn{2}{c}{r$_{c}$} &
\multicolumn{2}{c}{position ($\%$ of R$_{200}$)} \\
\multicolumn{1}{c}{r$_{c,min}$} &
\multicolumn{1}{c}{r$_{c,max}$ (Mpc)} &
\multicolumn{1}{c}{$Q_1$} & \multicolumn{1}{c}{$Q_3$} &
\multicolumn{1}{c}{$Q_1$} & \multicolumn{1}{c}{$Q_3$} &
\multicolumn{1}{c}{$Q_1$} & \multicolumn{1}{c}{$Q_3$}
}
\startdata
\nodata & 0.1 & 45.7$\%$ & 54.5$\%$ & 42.2$\%$ & 71.2$\%$ & 11.6$\%$ & 18.2$\%$  \\
0.1 & 0.2 & 43.3$\%$ & 53.4$\%$ & 45.1$\%$ & 87.3$\%$ & 15.0$\%$ & 25.8$\%$ \\
0.2 & 0.3 & 39.9$\%$ & 51.0$\%$ & 46.9$\%$ & 106$\%$ & 18.1$\%$ & 32.1$\%$ \\
0.3 & 0.4 & 37.5$\%$ & 48.9$\%$ & 44.4$\%$ & 89.7$\%$ & 18.5$\%$ & 33.0$\%$ \\
0.4 & 0.5 & 35.6$\%$ & 47.0$\%$ & 36.9$\%$ & 70.4$\%$ & 17.5$\%$ & 33.2$\%$ \\
0.5 & \nodata & 32.2$\%$ & 43.9$\%$ & 20.2$\%$ & 43.5$\%$ & 15.7$\%$ & 31.6$\%$\\
\enddata
\tablecomments{Ranges are from r$_{c,min} <$ r$_{c} \le$ r$_{c,max}$
with the same constraints on $\Lambda_{200}$ and r$_{c}$ as
table \ref{rh_table}.
$Q_1$ and $Q_3$ are the lower and upper quartiles, respectively.}
\end{deluxetable}

\begin{deluxetable}{llcccccc}
\tabletypesize{\footnotesize}
\tablewidth{340.0pt}
\tablecolumns{7}
\tablecaption{Errors in cluster redshift estimates\label{zpzs_table}}
\tablehead{
\multicolumn{1}{c}{} & \multicolumn{1}{c}{} &
\multicolumn{1}{c}{average offset} & 
\multicolumn{1}{c}{$\sigma_{z_{cl} - z_p}$} &
\multicolumn{1}{c}{$\%$ of clusters with} & 
\multicolumn{1}{c}{$\%$ of clusters with} &
\multicolumn{1}{c}{number of clusters}\\
\multicolumn{1}{c}{z$_{min}$} &
\multicolumn{1}{c}{z$_{max}$} &
\multicolumn{1}{c}{($z_{cl} - z_p$)} & \multicolumn{1}{c}{} &
\multicolumn{1}{c}{$|z_{cl} - z_p| > 3 \sigma_{z_{cl} - z_p}$} &
\multicolumn{1}{c}{$|z_{cl} - z_s| > \sigma_{z_{cl} - z_p}$} &
\multicolumn{1}{c}{in sample}
}
\startdata
\nodata & 0.1 & 0.0068 & 0.0070 & 2.4$\%$ & 0.0$\%$ & 543 \\
0.1 & 0.2 & 0.0010 & 0.0074 & 1.2$\%$ & 0.0$\%$ & 1549 \\
0.2 & 0.3 & -0.0003 & 0.013 & 1.0$\%$ & 1.7$\%$ & 3528 \\
0.3 & 0.4 & 0.0005 & 0.014 & 0.9$\%$ & 1.3$\%$ & 5016 \\
0.4 & 0.5 & -0.0009 & 0.013 & 0.5$\%$ & 1.6$\%$ & 3533 \\
0.5 & \nodata & 0.0019 & 0.022 & 1.3$\%$ & 2.9$\%$ & 557 \\
total & \nodata & 0.0003 & 0.013 & 0.9$\%$ & 1.1$\%$ & 14746 \\
\enddata
\tablecomments{Ranges are from z$_{min} <$ z $\le$ z$_{max}$;
clusters included in the sample have either five or more
spectroscopic redshifts or, for clusters with z $>$ 0.2, 
an associated BCG candidate with
a spectroscopic redshift measurement.}
\end{deluxetable}

\subsection{BCGs in the New Catalog}

Since different workers define what they mean by the Brightest
Cluster Galaxy somewhat differently, it is not surprising that
the assignment of the BCG in a given cluster is not always
the same.
For this reason
we decided to provide a list of the three galaxies that are
brightest in the $r$-band which belong to a specific cluster
to aid in comparisons with other catalogs.  This list omits
galaxies that have no redshift estimate, as those galaxies
have a likelihood of being present due to projection effects.  
The brightest
of the candidates is indicated in the list (see Appendix).

We binned 
the $\Lambda_{200}$ values for the BCG candidates as shown
in table \ref{bcg_mr} and computed the mean and standard deviation
of M$^{r}$ for both the brightest BCG candidate and for all
3 candidates together.  Using M$^{r}$~=~-2.5logL$_{r}$, where
L$_{r}$ is the luminosity in the $r$-band in terms of L$^{*}$,
we fit $L_{r}=\beta \Lambda_{200}^{\alpha}$ and get 
$\alpha$~=~0.18$\pm$0.01 and $\beta$~=~8.4$\pm$0.1.

\begin{figure}
\plotone{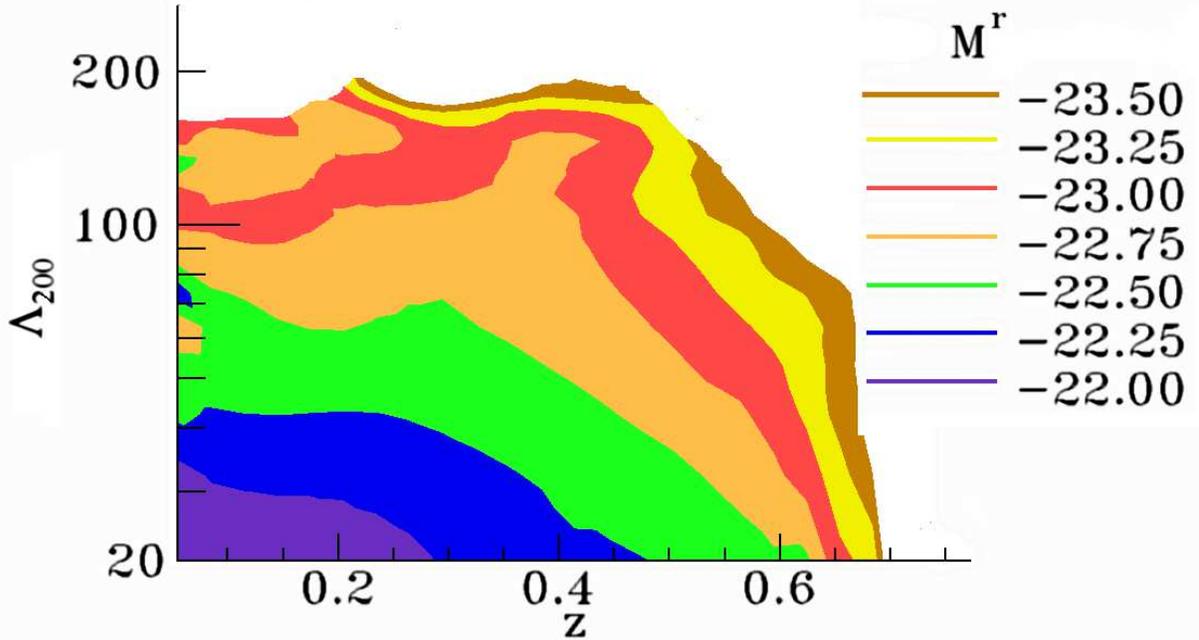}
\caption{Average absolute $r$-band magnitude vs. redshift and 
$\Lambda_{200}$ for the brightest BCG for each cluster.
Each contour marks the minimum value of M$^{r}$ for
that region.  The magnitudes presented are k-corrected based
on the values of z in the AMF catalog using k-corrections
as described in \citet{Blanton03}.
\label{bcgcontour}}
\end{figure}

The evolution of the luminosity 
of BCGs with respect to redshift has been discussed in numerous
papers, most of which assume that the luminosity of the
BCG will change with redshift in the same manner that M$^{*}$ does
\citep{linmohr04,zheng05,hansen05,dong08}.
We examine the evolution of M$^r$ for our brightest 
BCGs in both redshift and richness in 
Figure \ref{bcgcontour}.  We note that there is an evolution
of M$^r$ for any given redshift and any given richness.  We also
note that the behavior of M$^r$ is flat to a higher redshift
for higher richnesses for $\Lambda_{200} <$ 100.  Thus, for
clusters with $\Lambda_{200}$=25, the average luminosity of the BCG
is flat to $z \sim$0.2, whereas for clusters with 
$\Lambda_{200}$=50, the average luminosity is flat to $z \sim$0.4.
This trend indicates that BCGs in larger clusters formed
earlier than BCGs in smaller clusters, thus the more rapid
portion of their evolution has ended before z=0.6.
For clusters with $\Lambda_{200} >$100 or $z >$0.6, there are
not enough clusters to determine whether the trends established
at lower redshifts and richnesses continue. 

\begin{figure}
\plotone{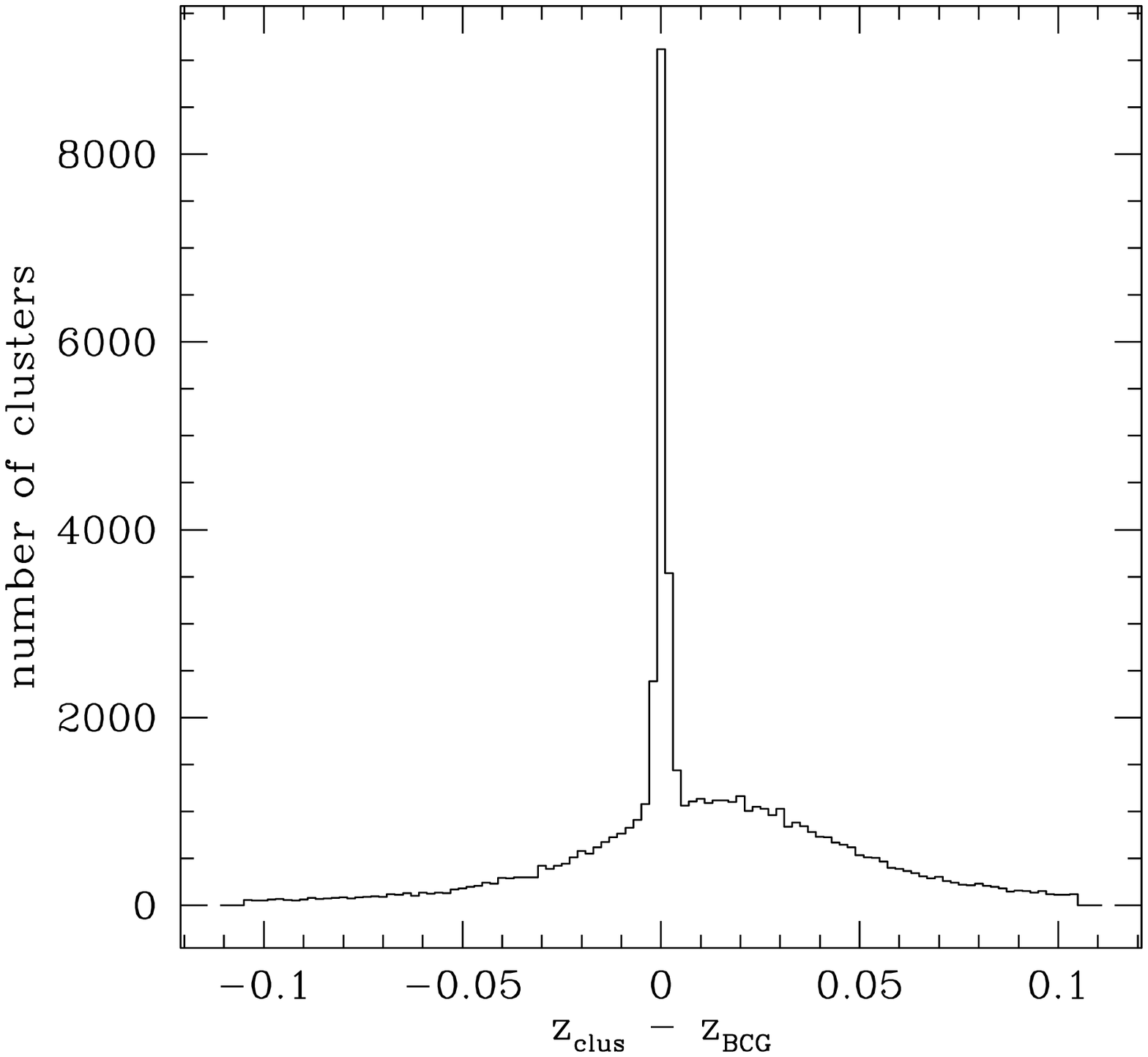}
\caption{Distribution of the difference between cluster redshifts and 
the redshift of the assigned brightest galaxy; the sharp peak at zero
is for BCGs with spectroscopic measurements, and the broader Gaussian
is due to BCGs with photometric redshift estimates only.  BCGs with
no redshift information have been excluded. \label{bcg_clus_dz}}
\end{figure}

We also examine the distributions
of the difference in redshift between the cluster and the
galaxy (Fig. \ref{bcg_clus_dz}).
In redshift, the distribution shows a large
peak near zero for galaxies with a spectroscopic redshift
and a broader Gaussian for those with only photometric 
redshifts.  We find that for the brightest galaxies,
there is a bias that assigns photometric redshifts that
are too large compared with spectroscopic redshifts for
the same galaxy.  This bias is reflected in the broad 
peak of the photometric redshifts in Fig. \ref{bcg_clus_dz}
being shifted by $\Delta$z~=~0.015 compared to the
spectroscopic peak.  
  
\begin{deluxetable}{llcc}
\tabletypesize{\footnotesize}
\tablewidth{230.0pt}
\tablecolumns{4}
\tablecaption{Average M$_{r}$ as a function of $\Lambda_{200}$\label{bcg_mr}}
\tablehead{\colhead{$\Lambda_{200,min}$} &
\colhead{$\Lambda_{200,max}$} &
\colhead{M$_{r}$ (brightest BCG)} &
\colhead{M$_{r}$ (3 BCGs)}
}
\startdata
120 & \nodata & -23.0$\pm$0.7 & -22.6$\pm$0.6 \\
100 & 120 & -22.9$\pm$0.6 & -22.5$\pm$0.6 \\
80 & 100 & -22.8$\pm$0.7 & -22.4$\pm$0.6 \\
60 & 80 & -22.7$\pm$0.6 & -22.4$\pm$0.7 \\
50 & 60 & -22.6$\pm$0.8 & -22.3$\pm$0.6 \\
40 & 50 & -22.6$\pm$0.7 & -22.2$\pm$0.6 \\
30 & 40 & -22.4$\pm$0.7 & -22.0$\pm$0.6 \\
20 & 30 & -22.3$\pm$0.7 & -21.9$\pm$0.7 \\ 
\enddata
\tablecomments{Richness ranges are determined such that
$\Lambda_{200,min} \le  \Lambda_{200} < \Lambda_{200,max}$.}
\end{deluxetable}

\section{Comparisons with Other Catalogs \label{comp_sect}}

In order to compare properties such as richness for two different
catalogs, it is necessary to determine one-to-one matches for
clusters between those catalogs.  
We match clusters in the AMF catalog with clusters from other catalogs 
by searching clusters within a given radius and 
redshift from each AMF cluster center.
As for the searching radius, we adopt 
the AMF R$_{200}$ value of the cluster in hand when 
comparing with maxBCG, WHL, and GMBCG catalogs, 
and 1.5 h$^{-1}$ Mpc when comparing with X--ray data; 
while we allow for a maximum redshift difference of 0.05.
In most cases, clusters in both the AMF catalog and a second
catalog are solitary, and there is no need for additional 
matching criteria.  The matches are unambiguous.  For the
remaining cases, the AMF finder determines several clusters
where only one is listed in the other catalog, or, more 
frequently, the other catalog determines multiple clusters
where the AMF catalog lists just one. 
In these instances, we refine the matching according to the procedure 
outlined in Appendix \ref{match_app} in order to 
obtain a one--to--one matching between the two catalogs.

One-to-one matches occur in 66$\%$ of potential matches
between the AMF catalog and the maxBCG catalog, 86$\%$ of
potential matches between AMF and WHL, 
76$\%$ of potential matches between AMF and GMBCG,
and 60$\%$ of 
potential matches between AMF and our X--ray cluster sample.
Additionally, a comparison of the AMF catalog and the 
Abell catalog finds one-to-one matches for 90\% of the
AMF clusters in the North Galactic Cap region (stripes 9-39
and 42-44).  The minimum $z$ for the AMF finder is 0.045,
whereas there is no minimum for the Abell catalog.  Of the
1179 AMF clusters matched in the Abell catalog, we measure
$z <$~0.25 for 861 (73$\%$) of them. 

\subsection{Comparison with maxBCG \label{maxbcg_clus_comp}}

The most widely used SDSS catalog to date is the maxBCG catalog compiled by
\citet{koester07}. MaxBCG  extends 
over a $\sim$7,500 deg$^{2}$ of the $\sim$8,420 deg$^{2}$
area covered by the AMF catalog and is restricted to $0.1 \le z \le 0.3$.
In this region, there are 12,761 AMF
clusters between z=0.1 and z=0.3, and 16,756 AMF clusters
between z=0.08 and z=0.33 (i.e. clusters that might enter the sample due
to errors in redshift).  The matching procedure
defined at the beginning of this section finds 5,447 matches.

Furthermore, using the fact that the given position of a
maxBCG cluster is the position of the BCG for that cluster, 
we identified maxBCG clusters for which the
quoted BCG correspond to one of our three brightest galaxies.
For unique pairings with cluster redshifts within
0.05 of each other, we found 4317 such BCG matches. 
We studied the properties of the ~1000 maxBCG BCGs 
which have a geometric match with one of our clusters
but do not appear in our BCG list.  We found that 
 they are systematically fainter than the AMF quoted BCG,
or that one of the AMF BCGs deviates by more than 0.2 mag
from the red-sequence.

\begin{figure}
\plotone{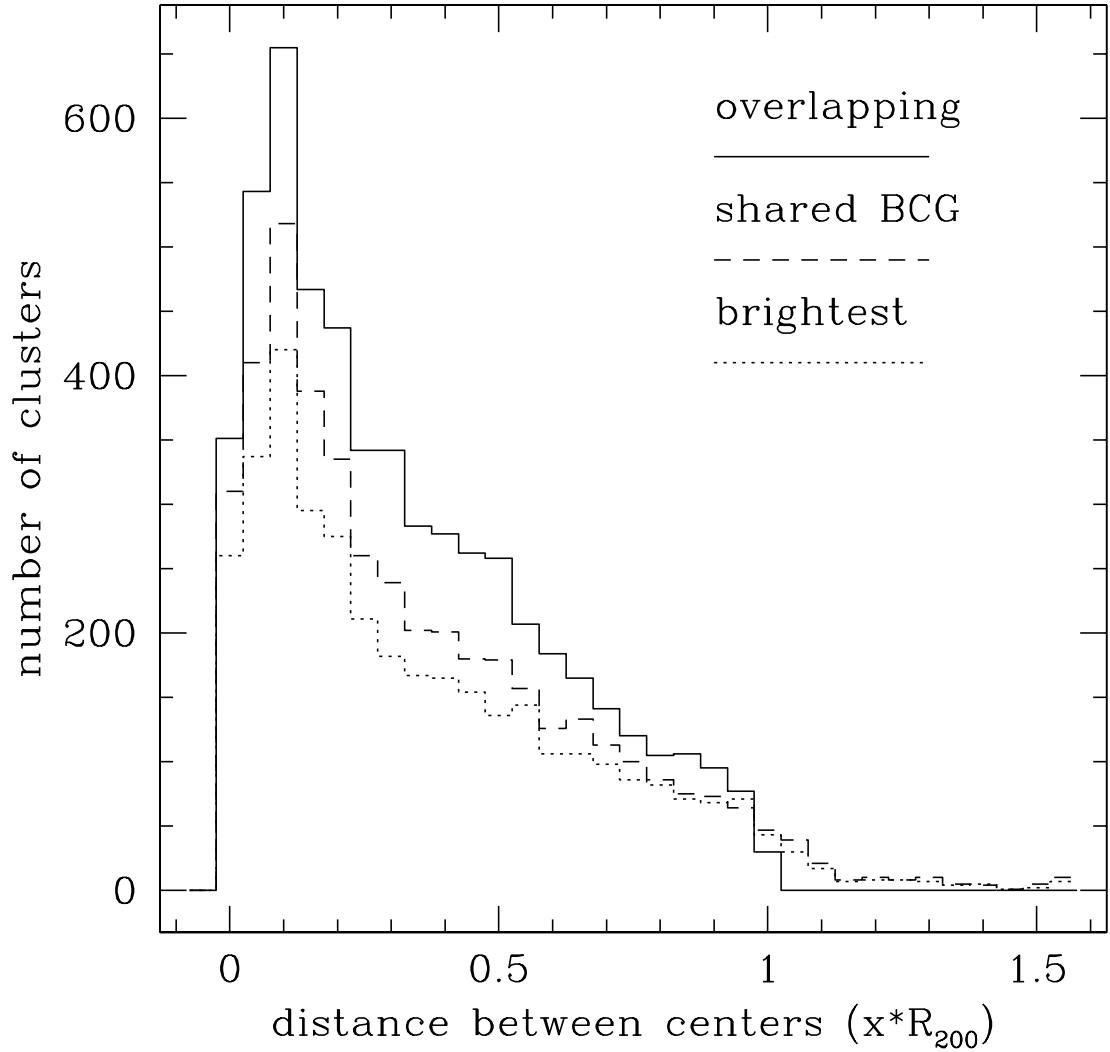}
\caption{Separation between AMF and maxBCG cluster centers
for clusters whose centers are within R$_{200}$ of the
AMF cluster (solid line), those which also share a BCG 
candidate from our catalog (dashed line), and those where
our brightest BCG candidate matches the maxBCG BCG (dotted line).
\label{amf_maxbcg_fsep}}
\end{figure}

The separation
between centers as a fraction of the AMF cluster R$_{200}$
value is shown in figure \ref{amf_maxbcg_fsep}, 
showing that a significant fraction of AMF clusters 
have centers quite distant from the quoted maxBCG one.
The histograms for the shared BCG candidates and brightest
BCGs keep the same proportion with the curve for those 
clusters that overlap.

\begin{figure}
\plotone{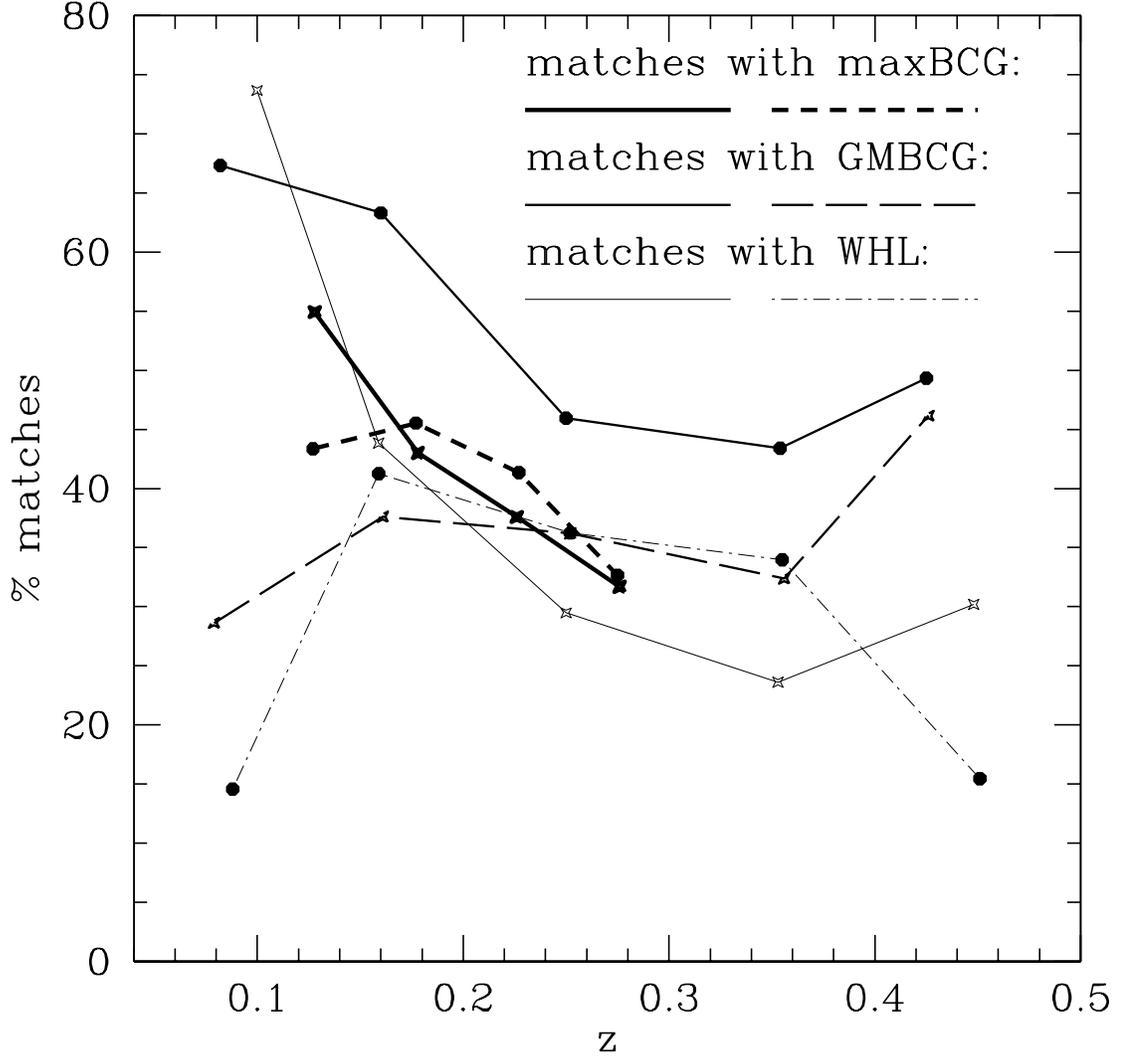}
\caption{Percentage matches between the AMF catalog and other
catalogs by redshift bin; 
the solid line
represents the percentage of clusters from the other catalog matched with
an AMF cluster, and the segmented line shows the percentage
of AMF clusters matched with with a 
cluster in the other catalog. The heaviest line and
dotted line represent maxBCG matches. The medium
line and dashed line represent GMBCG matches.
The thinnest line and dashed-dotted line represent
WHL matches. \label{z_pct_other}}
\end{figure}

\begin{figure}
\plotone{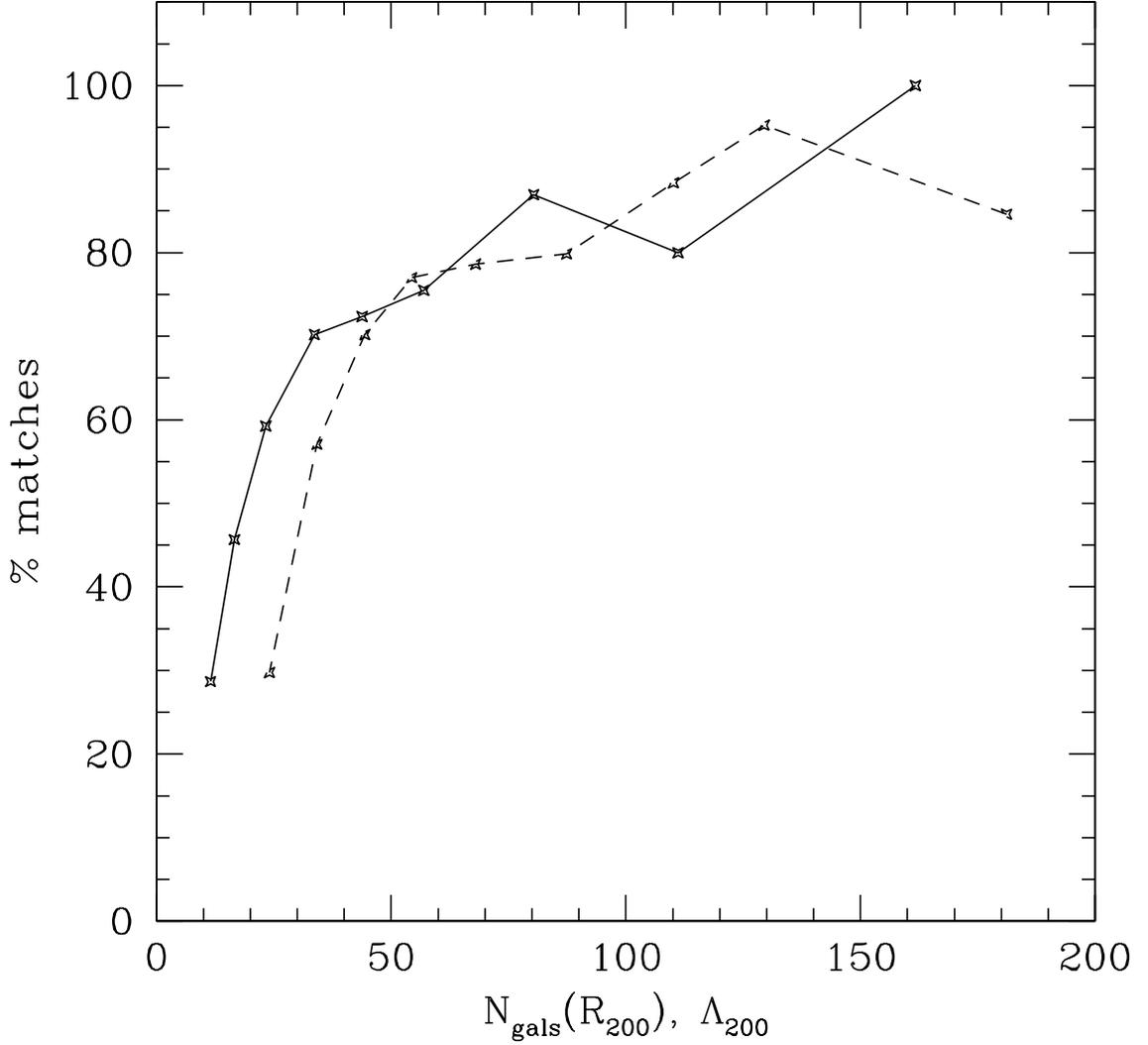}
\caption{Percentage of the matches between the AMF catalog and
the maxBCG catalog as a function of richness; the solid line
shows the
percentage of maxBCG clusters that have an AMF match as a 
function of $N_{gals}$($R_{200}$); the dashed line
indicates the percentage
of AMF clusters in the 0.1 $\le$ z $\le$ 0.3 range with a 
maxBCG match as a function of $\Lambda_{200}$.\label{maxbcg_rh_comp}}
\end{figure}

Figure \ref{z_pct_other} shows
the percentages of clusters from one catalog that are
matched in the other catalog as a function of redshift, and
figure \ref{maxbcg_rh_comp} shows this as a 
function of richness for matches with the maxBCG catalog.
Both of the percentage comparisons are based on the
5,447 overlapping cluster matches.
These figures show that the AMF catalog recovers 
80$\%$ of all maxBCG clusters above $N_{gal} = 80$,
but decreases detection to 30\% for the smallest 
maxBCG quoted richness ($N_{gal} = 10$).
Similarly, maxBCG finds about 75\% of AMF 
clusters above $\Lambda_{200}=50$ but drops to
30\% detections for $\Lambda_{200}=20$.

These differences should not be surprising, as   
the richness criteria for the two catalogs are different.
 While the AMF catalog relies on the total
luminosity within R$_{200}$, the maxBCG catalog counts
any collection of ten or more galaxies that satisfy a
luminosity cutoff and color relation as a cluster. 
The minimum luminosity within R$_{200}$ for one of
maxBCG clusters is potentially as 
small as 4L$^{*}$  In addition, \citet{koester07} uses
$i$-band values for their luminosity criteria,
while we use $r$-band values.

\begin{figure}
\plotone{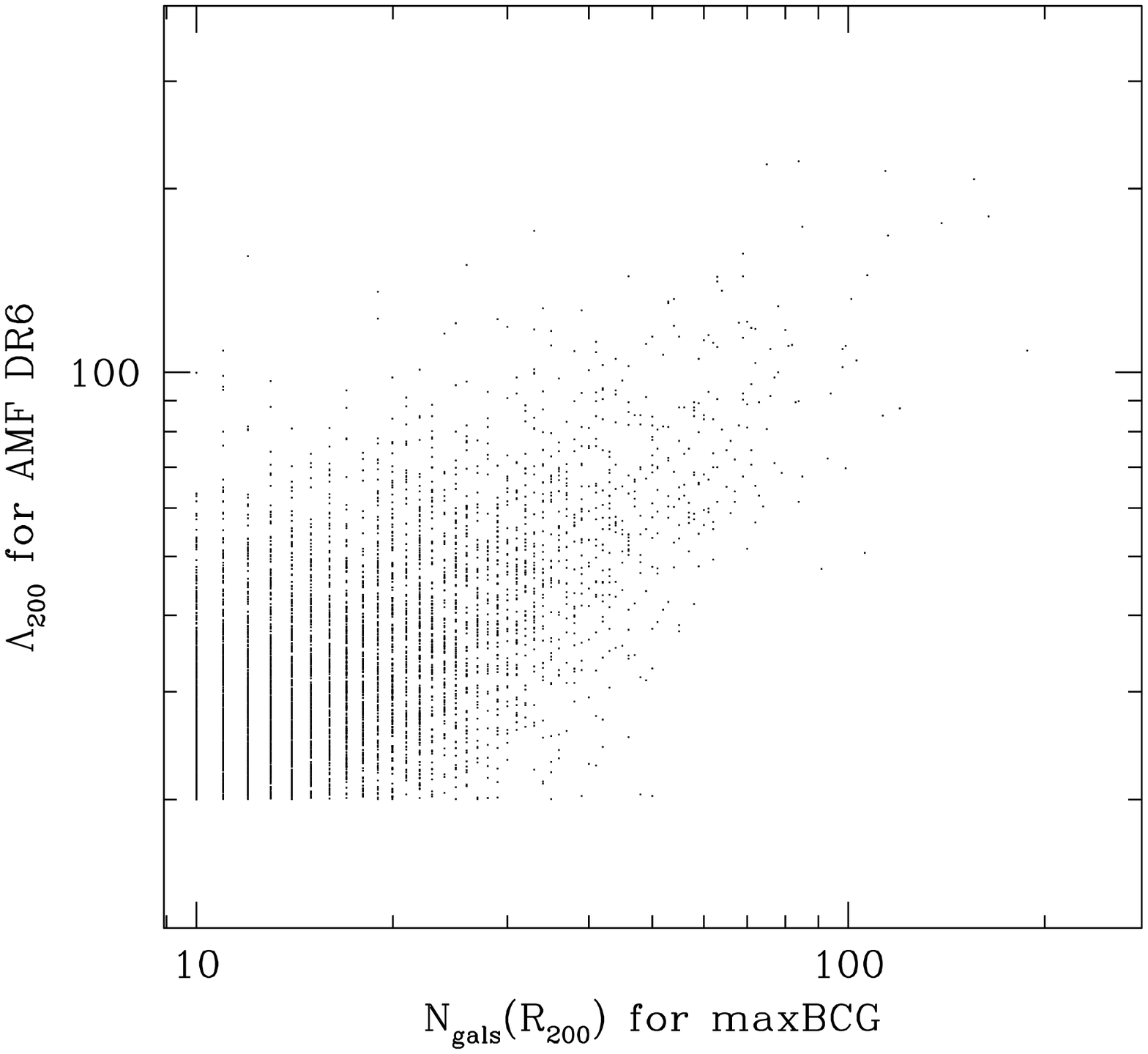}
\caption{$\Lambda_{200}$ vs. $N_{gals}(R_{200})$ for matching clusters.
\label{rh_ngal_comp}}
\end{figure}

To compare richnesses for matching clusters, we use
the N$_{gals}$(R$_{200}$) value to represent
clusters from \citet{koester07}.
Fig.~\ref{rh_ngal_comp} shows the distribution
of the richness measurements for each catalog for
these clusters using the sample of 5,447 matches.
For blended clusters, we always match the maxBCG cluster
to the AMF cluster at that site with the highest
$\Lambda_{200}$ value.  These are also the clusters
on a site with the highest likelihood value. 
A relationship between the richness values does not
become evident until N$_{gals}$(R$_{200}$)~$\gtrsim$~20.
At such richness, the matches between the two catalogs
reaches the 50$\%$ level.  
Below this value, the  $\Lambda_{200}$ for a given 
maxBCG cluster may vary by an order of magnitude.
Using the errors on $\Lambda_{200}$ values as weighting
factors, a best fit line to Fig.~\ref{rh_ngal_comp} is
\begin{equation}
\Lambda_{200}=1.67 N_{gals}(R_{200}) + 31.3.
\end{equation}
The same analysis performed on clusters matched on their BCGs
rather than on angular proximity produces a nearly identical
relation.  For BCG matches, the error-weighted best fit is
\begin{equation}
\Lambda_{200}=1.61 N_{gals}(R_{200}) + 34.8.
\end{equation}
Note that in both cases the relations are only meaningful for
clusters with N$_{gals}$(R$_{200}$)~$\ge$~30 or 
$\Lambda_{200} \ge$~80.  The errors in $\Lambda_{200}$ are
significantly smaller for larger values of $\Lambda_{200}$ 
(see Section \ref{err_det_sect}), thus matches with 
high richness AMF clusters 
tend to dominate a linear fit.

A thorough comparison aimed at explaining the discrepancies 
between the two catalogs is quite difficult because not 
only the two searching methods differ, but also choices 
on data pre-processing differ.
Specific choices on photometric redshifts used, k-correction, 
band for BCG detection, method for pre--selection of the 
galaxies to include in the search are all factors that 
may play a role in the final output.
We therefore limited the analysis of the discrepancy  
to the study of luminosity and color properties of the
BCGs for matching clusters and compare them with our whole BCG sample.
We computed the offset from the red 
sequence and the absolute magnitude 
of BCG (according to the DR6 color frame) in the following BCG samples:
$i)$ maxBCG BCG for  clusters  in common between the two 
catalogs and found among the three AMF brightest galaxies, 
compared with the brightest AMF BCG
for the same clusters.
 In clusters that AMF and maxBCG share, the maxBCG BCG selection
coincides with that of the AMF catalog. Only in ~9\% of 
the cases maxBCG misses the 
brightest galaxy because it  is too blue. In such cases maxBCG 
uses the 2nd or the 3rd as seed for the cluster.
The fraction of blue BCG found here is consistent with what 
is found for rich systems ($\Lambda_{200} > 50$) in the whole 
AMF catalog (6.2$\%$, see \citet{Pipino10}, fig. 2).
$ii)$ maxBCG BCG for clusters  in common between the two 
catalogs, compared with AMF
first ranked BCG for all AMF clusters below  $z\le 0.3$.
An increase in the tail of blue BCG in the AMF catalog 
is apparent, now amounting
to 15.8\% of the cases. Note that 14.6\% is the fraction of 
blue first ranked galaxies in  all AMF clusters.
AMF BCGs are also slightly fainter than maxBCG ones, 
because here we are including in the comparison several  
clusters that don't match with maxBCG ones.
$iii)$ first ranked AMF BCGs in the redshift range 
$0.1\le z \le 0.3$ that are not in clusters
that share the BCG with maxBCG.
The AMF BCGs in this case are on average fainter 
(as we tend to retain here  only poor systems) and the 
fraction of blue BCG goes up to 16$\%$.
The existence of such blue BCGs in our sample is intriguing, 
and it would be important to further confirm it with spectroscopic studies
of this sample.

While the differences in colors between the AMF and 
maxBCG BCGs are evident, they don't seem 
to be sufficient to explain the quite substantial 
differences between the two catalogs.
 It is possible  that the biggest difference in the low 
luminosity end of the two catalogs  is due to maxBCG
requiring clusters to have at least ten ridgeline galaxies 
found in the proximity of the selected BCG. These  
color cuts reduce the number of galaxies in the input 
SDSS photometric database to one fifth with respect to the initial value.

A search for max BCG clusters in an extended AMF catalog 
also retaining systems with richness  in the 
range $12 \le \Lambda_{200} < 20$  leads to a 33$\%$ 
increase in the number of matches, from 6000 to 8000 matches. 
However, the completeness of the cluster finder is
$\sim$50$\%$ for this size cluster (D08).  Thus, it is possible that this 
increase of 33$\%$ represents only half of the matches gained by
lowering the minimum $\Lambda_{200}$ value, {\it i.e.} the
actual gain is 66$\%$, or 10,000 matches.  As we lower the
threshold further, the completeness decreases, making 
comparisons even more difficult.  

\citet{koester07} report the total luminosity in the $r$-band
of the members of their clusters as $L_{r}^{memb}$.
By accounting for the evolution of L$^{*}$ with redshift
for our clusters, we can make a direct comparison of
our $\Lambda_{200}$ values with their $L_{r}^{memb}$ values.
The range for 20L$^{*}$ for 
$z$ between 0.1 and 0.3 for our catalog is 
38 to 50 $\times$ 10$^{10}$ L$_{\sun}$.  There are
only $\sim$1600 maxBCG clusters with 
$L_{r}^{memb} \ge$ 40 $\times$ 10$^{10}$ L$_{\sun}$.
Because we know we match at least 5000 of their
clusters based on BCG, our clusters match
with maxBCG clusters whose luminosity in the $r$-band
is less than our richness cutoff of 20L$^{*}$.
A large percentage of maxBCG clusters are likely
to be too poor to be detected by our finder.

\subsection{Comparison with WHL Clusters}

\begin{figure}
\plotone{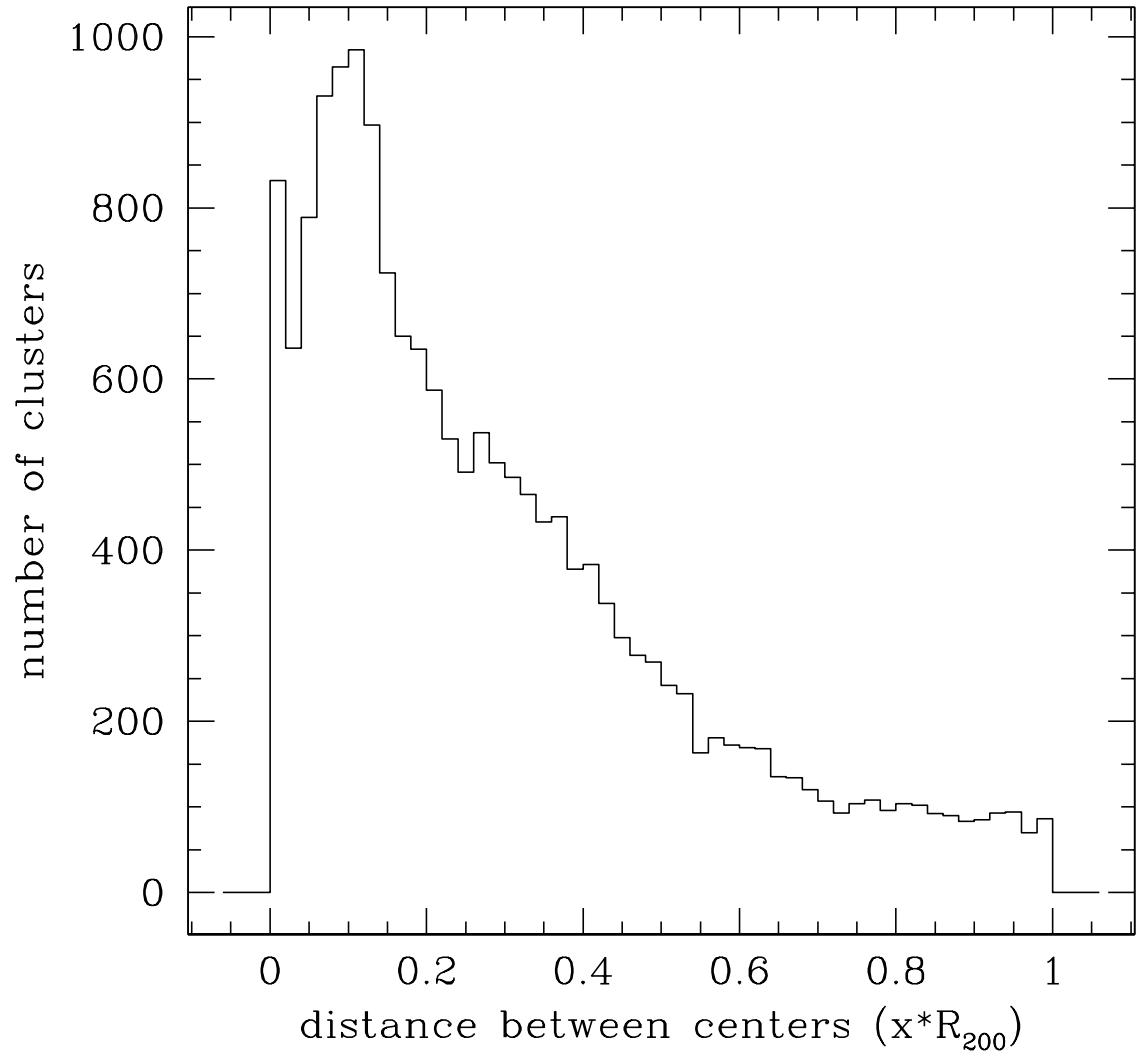}
\caption{Distribution of separations between the AMF cluster
centers and WHL BCGs for matching clusters.  \label{whl_sep}}
\end{figure}

\begin{figure}
\plotone{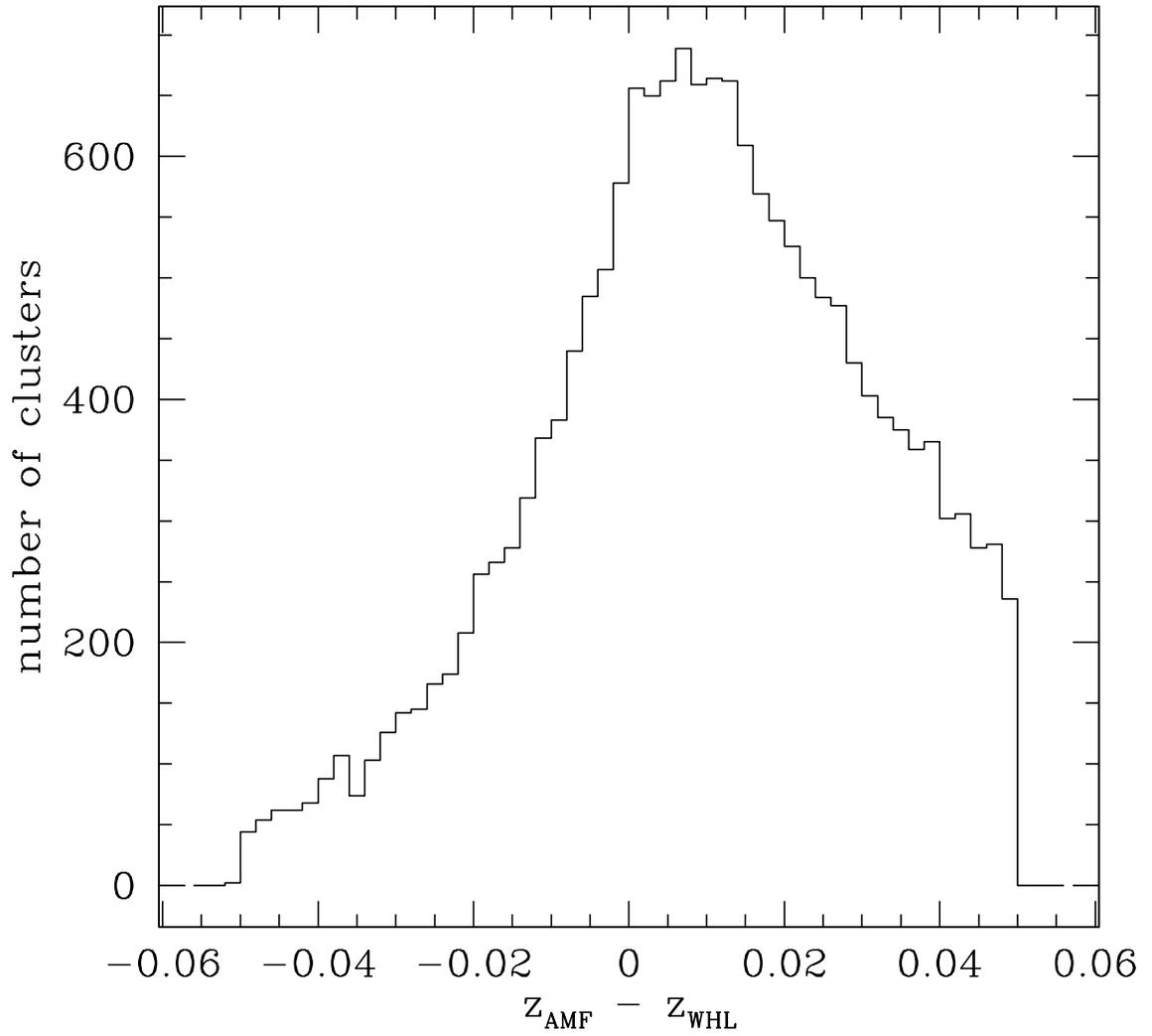}
\caption{Distribution of $\Delta z$ values for matching
AMF and WHL clusters.  \label{whl_z}}
\end{figure}

\begin{figure}
\plotone{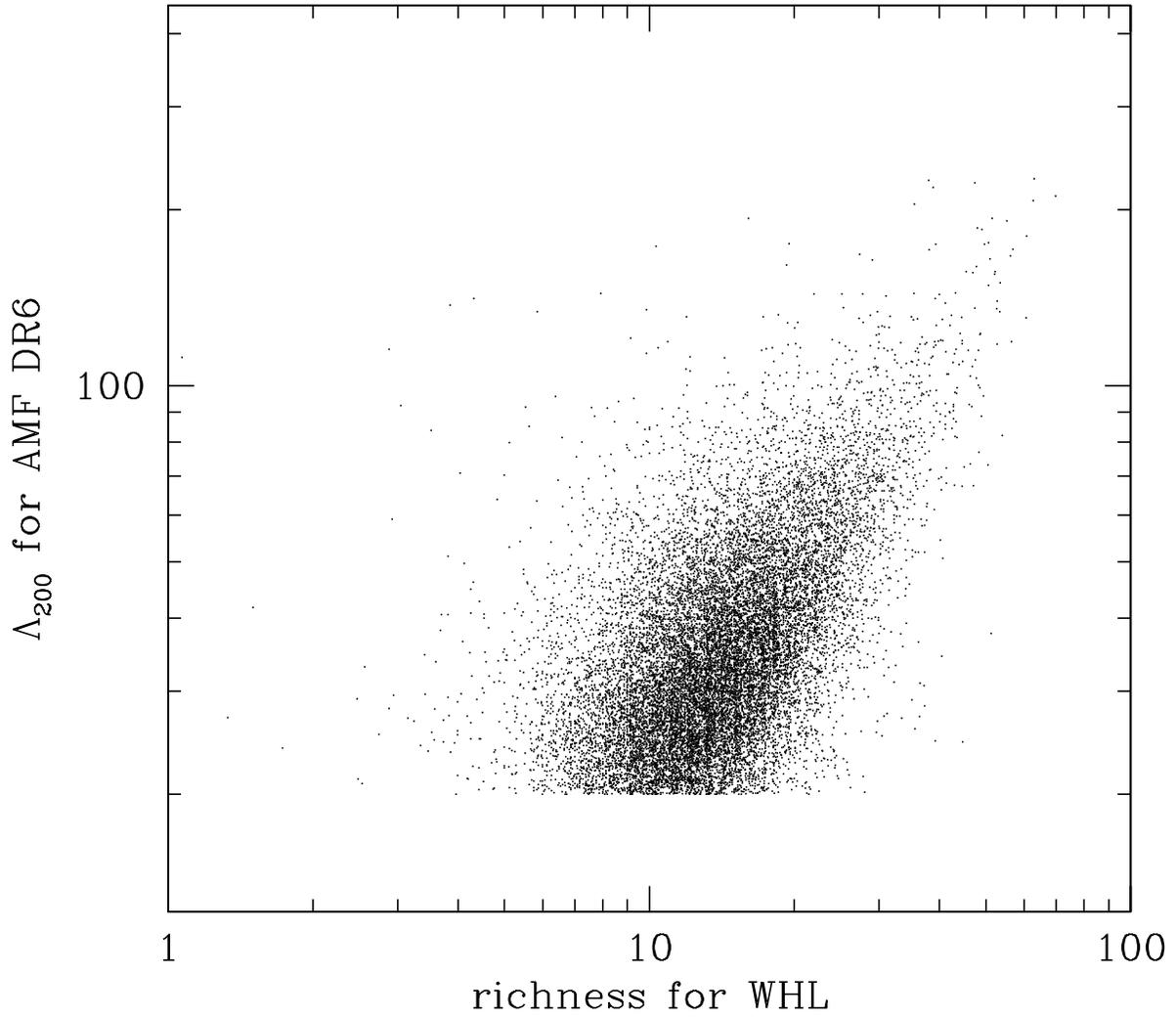}
\caption{$\Lambda_{200}$ values (AMF) vs. richness values (WHL) for
paired clusters.  \label{whl_rich}}
\end{figure}

The WHL catalog \citep{WHL09} was produced from the same
SDSS Data Release 
as our catalog.  The angular positions provided by \citet{WHL09} are of
the BCGs of their clusters.  The following comparisons are for
cases where the WHL BCG lies within our value of R$_{200}$ from
the center of our cluster, and $|\Delta z|$ between the 
photometric redshift of their cluster and the redshift
of our cluster is $\le$~0.05.  We find 17,579 pairs, or matches
for slightly less than 50$\%$ of their clusters.  As with the maxBCG
comparison, the method of cluster identification along with
the completeness of the surveys for smaller cluster sizes 
will affect the number of matches.  Figure \ref{whl_sep}
shows the distribution of separations of our cluster centers
and the WHL BCGs.  The peak in separations with this
binning occurs at a difference of 0.1 h$^{-1}$ Mpc --- the
same value as the comparison with maxBCG clusters. 
Figure \ref{whl_z} shows the
distribution of redshift differences between matching 
clusters.  There is a slight ($\Delta z = 0.006$) bias
toward AMF clusters being more distant; most paired clusters lie
within $|\Delta z| = 0.025$ of each other, which is to
be expected considering the error in redshift measurements
for each finding technique.  To compare sizes of paired
clusters, we examine how our $\Lambda_{200}$ value compares
with the richness reported by \citet{WHL09} in Figure \ref{whl_rich}.
The amount of scatter has decreased with respect to the
maxBCG comparison (Figure \ref{rh_ngal_comp}).  The 
error-weighted least-squares
fit to the data gives
\begin{equation}
\Lambda_{200} = 3.79 R + 13.7.
\end{equation}

\begin{figure}
\plotone{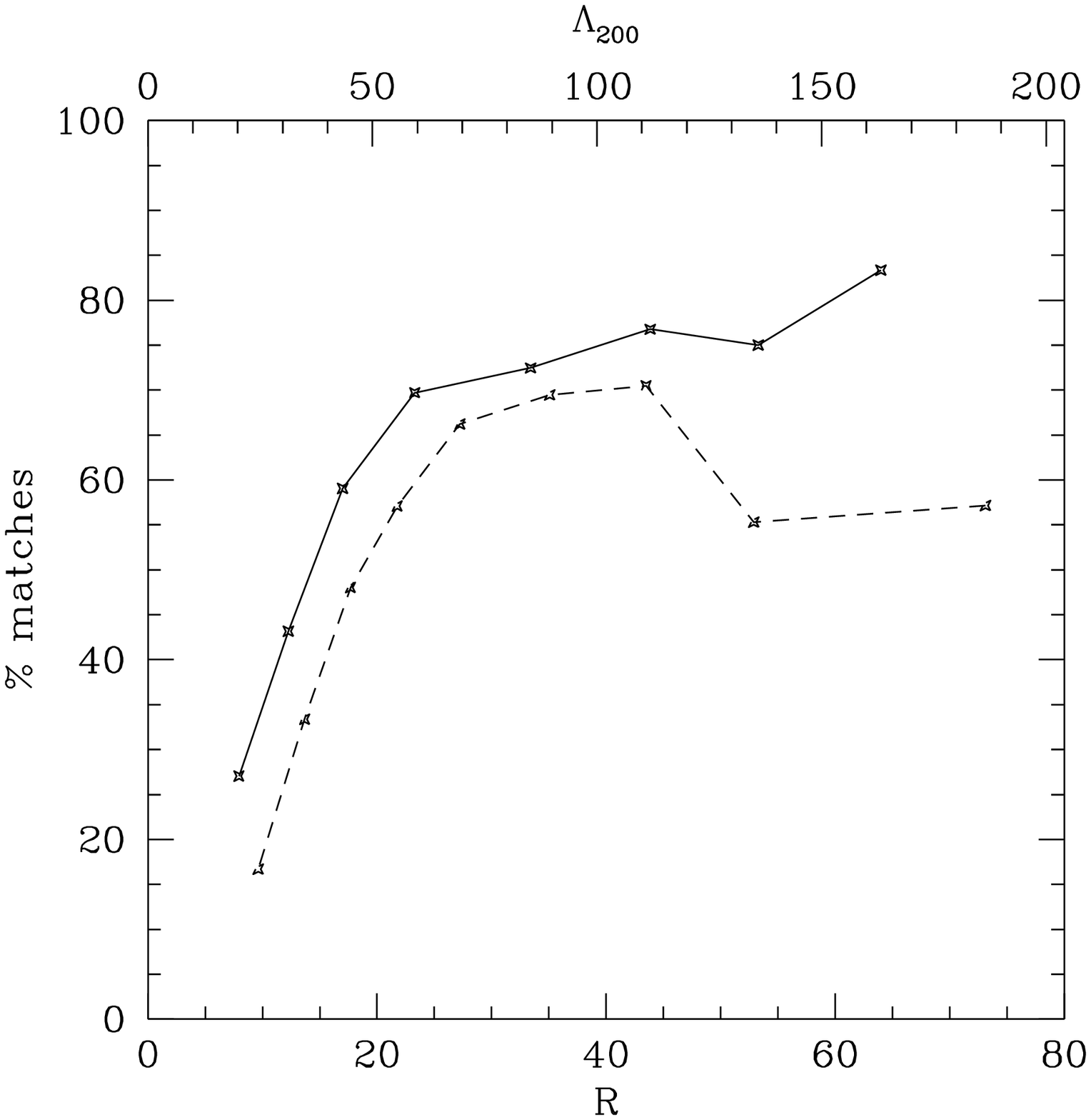}
\caption{Percentage of the matches between the AMF catalog and
the WHL catalog as a function of richness; the solid line
shows the
percentage of WHL clusters that have an AMF match as a 
function of $R$ (lower x-axis); the 
dashed line shows the percentage
of AMF clusters in the z $\le$ 0.6 range with a 
WHL match as a function of $\Lambda_{200}$ (upper x-axis).\label{whl_rh_comp}}
\end{figure}

In Figures \ref{z_pct_other} and \ref{whl_rh_comp}, we examine
the relative completeness of the AMF and WHL catalogs in 
redshift and richness with respect to each other.  The 
WHL catalog has matches in the AMF catalog for $>$70$\%$ of
its clusters with $R \ge$ 25.  The percentage of WHL clusters
with matches in our catalog increases to $>$80$\%$ for its
richest clusters.  The completeness of AMF clusters which
have matches in the WHL catalog follows a different trend.
For 20 $\le \Lambda_{200} \le $100, the percentage of matches
increases from $<$20$\%$ to just under 70$\%$.  Above
$\Lambda_{200}=$110, the percentage of matches falls to 
between 55 and 60$\%$.  In some cases, our larger clusters
are fragmented into several 
smaller clusters in the WHL catalog, and our lists
of matching pairs do not identify one-to-one matches in this case.
In redshift, Figure \ref{z_pct_other} shows that most of the
WHL clusters nearer than  $z=0.2$ have matches in the AMF
catalog ($>$60$\%$).  To the contrary, there are $<$40$\%$
matches in the WHL catalog for AMF clusters over the 
entire redshift range.  This can be attributed to the
fact that we find more clusters overall.

\subsection{Comparison with GMBCG}

The GMBCG catalog \citep{GMBCG} is the follow-up to the maxBCG catalog.
It employs a more sophisticated method for determining the magnitude
versus redshift distribution 
and extends the range from a maximum of $z =$ 0.3 (for maxBCG) to a maximum
of $z =$ 0.55.  This catalog still displays a richness distribution
similar to the maxBCG catalog in that it finds many objects that
are below the detection threshold of the AMF catalog.

There are 15,214 one-to-one matches between the AMF and GMBCG catalogs
in the region of the sky where the catalogs overlap and for $z \le $ 0.55.
The GMBCG catalog does not detect clusters in regions that were not
covered by DR6, nor does it include the portions of 
stripes 38 and 39 with RA $>$ 200$^{\circ}$ or stripe 44.
Figure \ref{gmbcg_z} shows the distribution in matching clusters
as a function of $z_{AMF} - z_{GMBCG}$.  More than 5,000 of the 
matching clusters have a $|\Delta z| \le $ 0.005.  

For one-to-one matching clusters, we plot the value
of $\Lambda_{200}$ vs. the value of N$_{gals}$ in
figure \ref{gmbcg_rich}.  We use the weighted value
of N$_{gals}$ if the {\tt WeightOK} flag is set,
otherwise we use the richness from {\tt GM\_Scaled\_Ngals}.
Using the errors on $\Lambda_{200}$ values as weighting
factors, a best fit line to Fig.~\ref{gmbcg_rich} is
\begin{equation}
\Lambda_{200}=1.37 N_{gals}(R_{200}) + 47.8.
\end{equation}
As the errors are smaller for richer AMF clusters, the 
fit is biased toward larger clusters.

Figure \ref{z_pct_other} shows the percentage of
clusters in one catalog that are matched in the other
as a function of redshift. 
Figure \ref{gmbcg_rh_comp} represents the percentage of the
AMF and GMBCG catalogs in richness which
have matches in the other catalog.
For $\Lambda_{200} > 80$ for AMF clusters and $N_{gals} > 55$
for GMBCG clusters, the catalogs have 60\% of their
members matched in the other catalog, except for the
highest richness clusters in the GMBCG catalog that match
more than 70\% of clusters in the AMF catalog.
The richest clusters in the AMF catalog are often fragmented
into multiple smaller clusters in the GMBCG catalog.  
The issue of clusters the GMBCG catalog having many
clusters below the richness threshold used for the
AMF catalog is the same issue we examined when comparing
our catalog to the maxBCG catalog in section \ref{maxbcg_clus_comp}.
This results
in many of our richest clusters not having a one-to-one match
with a GMBCG cluster.  
\begin{figure}
\plotone{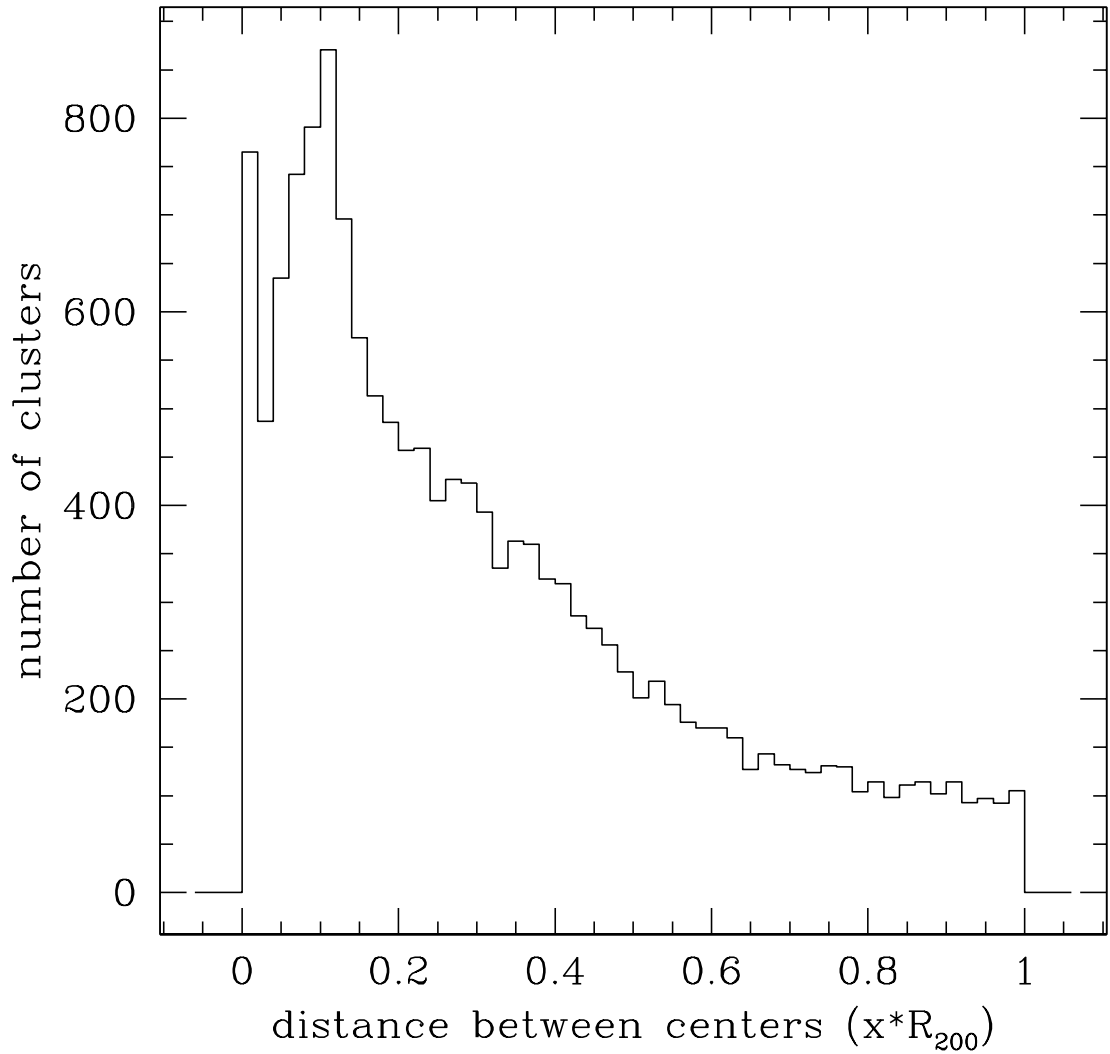}
\caption{Distribution of separations between the AMF cluster
centers and GMBCG BCGs for matching clusters.  \label{gmbcg_sep}}
\end{figure}

\begin{figure}
\plotone{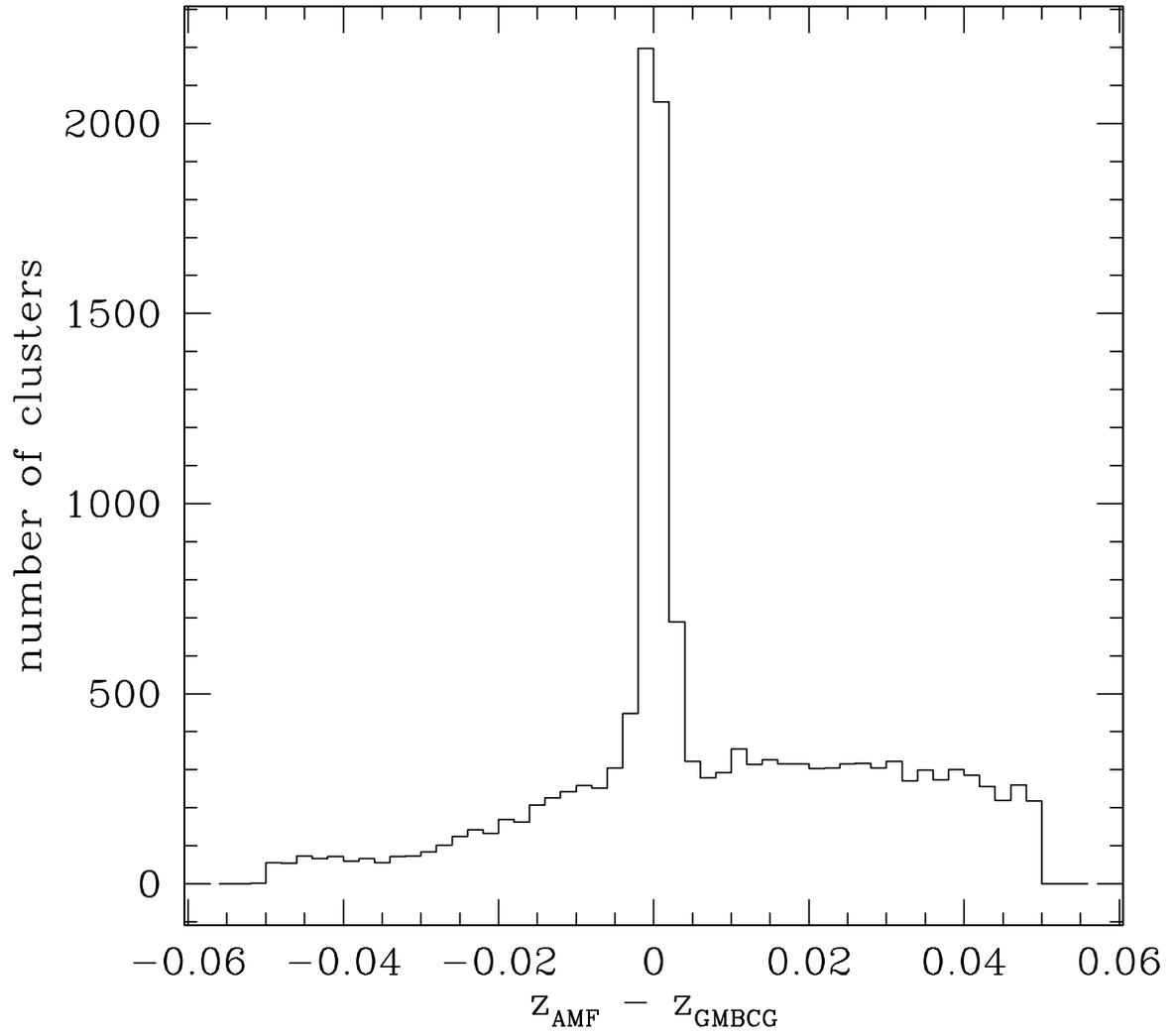}
\caption{Distribution of $\Delta z$ values for matching
AMF and GMBCG clusters.  \label{gmbcg_z}}
\end{figure}

\begin{figure}
\plotone{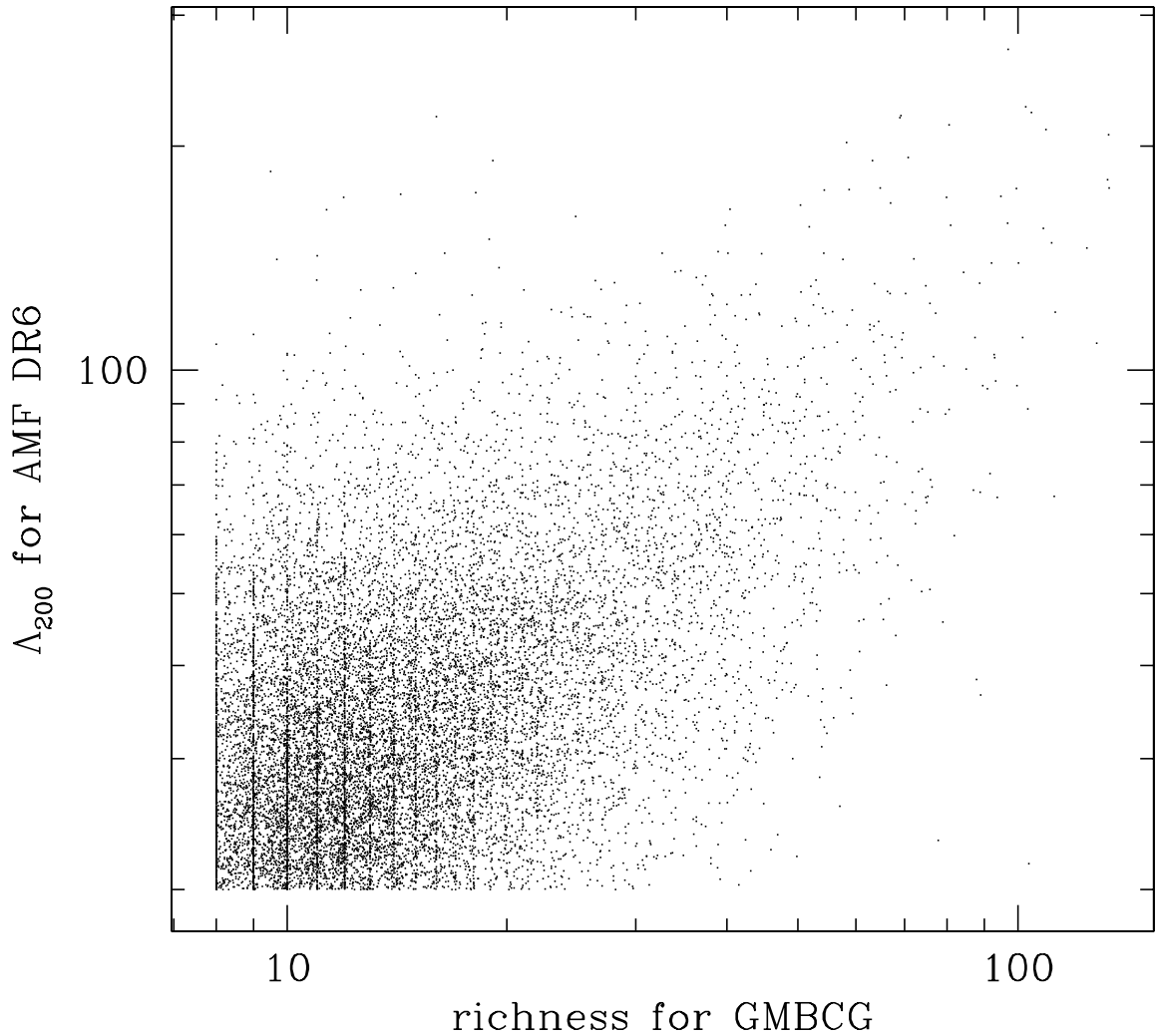}
\caption{$\Lambda_{200}$ values (AMF) vs. richness values (GMBCG) for
paired clusters.  The richness from the GMBCG catalog is
the weighted value of N$_{gals}$ if the {\tt WeightOK} flag
is set; elsewise, the scaled value of N$_{gals}$ is used. \label{gmbcg_rich}}
\end{figure}

\begin{figure}
\plotone{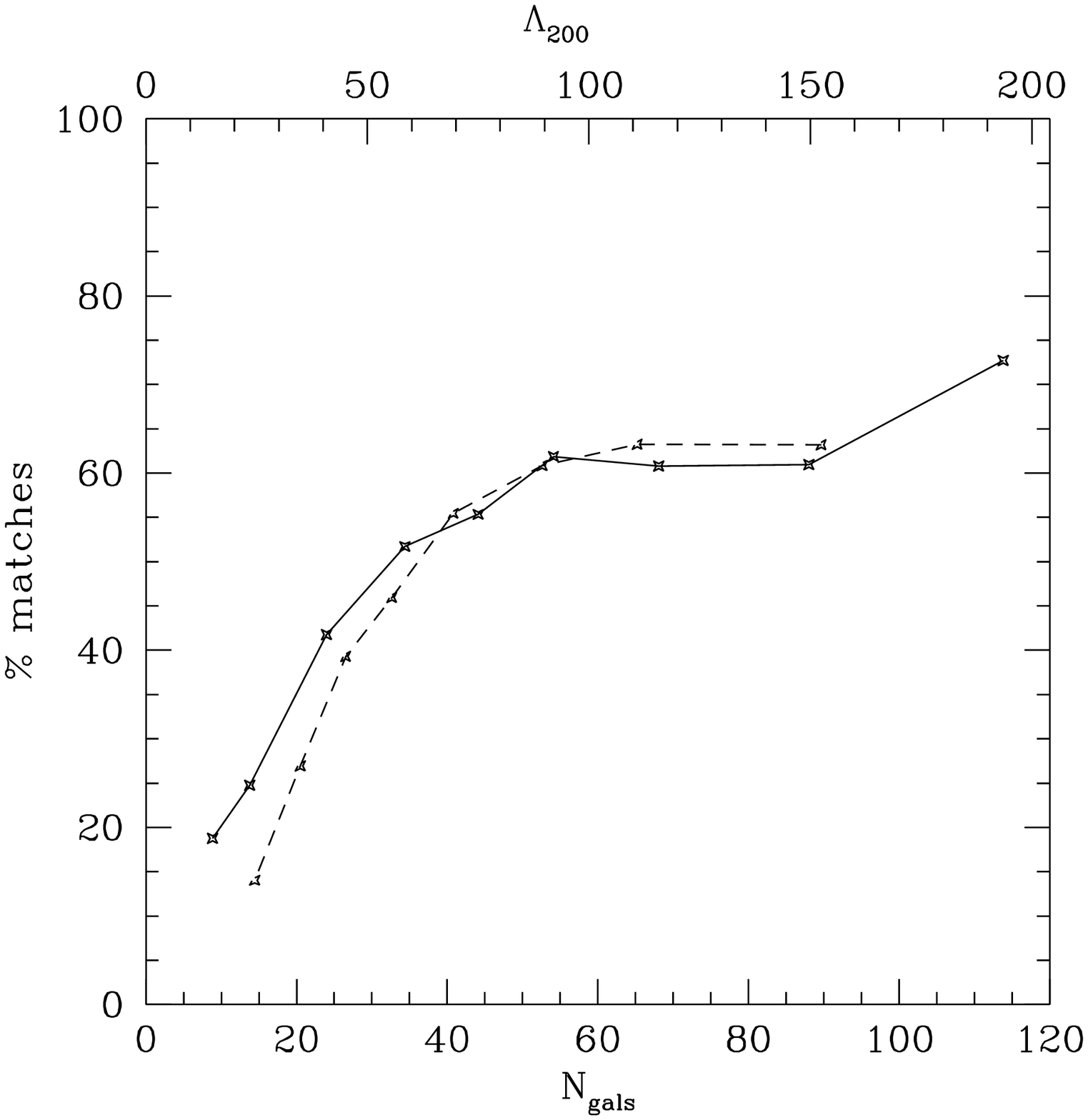}
\caption{Percentage of the matches between the AMF catalog and
the overlapping sections of the 
GMBCG catalog as a function of richness; the solid line
shows the
percentage of GMBCG clusters that have an AMF match as a 
function of
$N_{gals}$
 (lower x-axis); the
 dashed line shows the percentage
of AMF clusters in the z $\le$ 0.55 range with a 
GMBCG match as a function of $\Lambda_{200}$ 
(upper x-axis).\label{gmbcg_rh_comp}}
\end{figure}

\subsection{Comparison with X-ray Clusters}

In order to find a X-ray counterpart to  clusters in our catalog
we retrieved data from different sources and created an 
input list for a cross-matching procedure. The vast majority
of the clusters in such list comes from the X-ray Clusters Database (BAX)
as available on 24 June 2009. 
A BAX query returns nearly 1000 clusters and groups
in the region of the sky covered by DR6. Luminosities 
for the BAX objects have been recalculated in order 
to reflect the cosmology assumed in this paper.
Another 26 clusters from
very recent papers in the literature 
\citep{balestra06, maughan08, cavagnolo09} 
have also been added to the list. 
These works also provided (updated) temperature measurements.
While this compilation comprises the widest possible 
set of X--ray clusters we can match, it is not a 
well defined sample in terms of characteristics. We therefore 
also match our AMF sample with the flux--limited NORAS  
catalogs in the overlapping regions.

\begin{figure}
\plotone{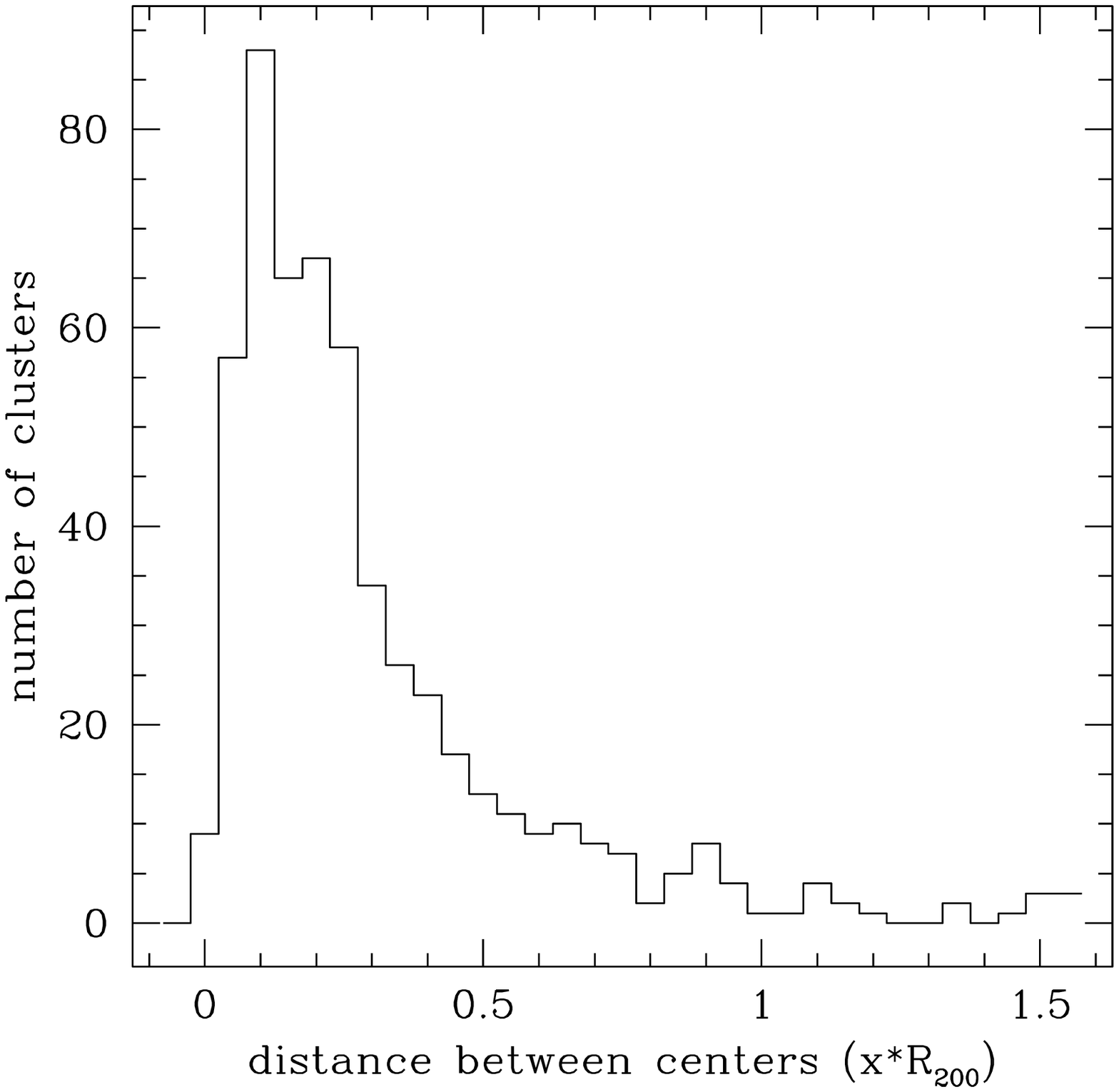}
\caption{Distance between the center of the AMF cluster and
the center of the BAX cluster as a fraction of 
R$_{200}$ of the AMF cluster.\label{bax_amf_sep}}
\end{figure}

\begin{figure}
\plotone{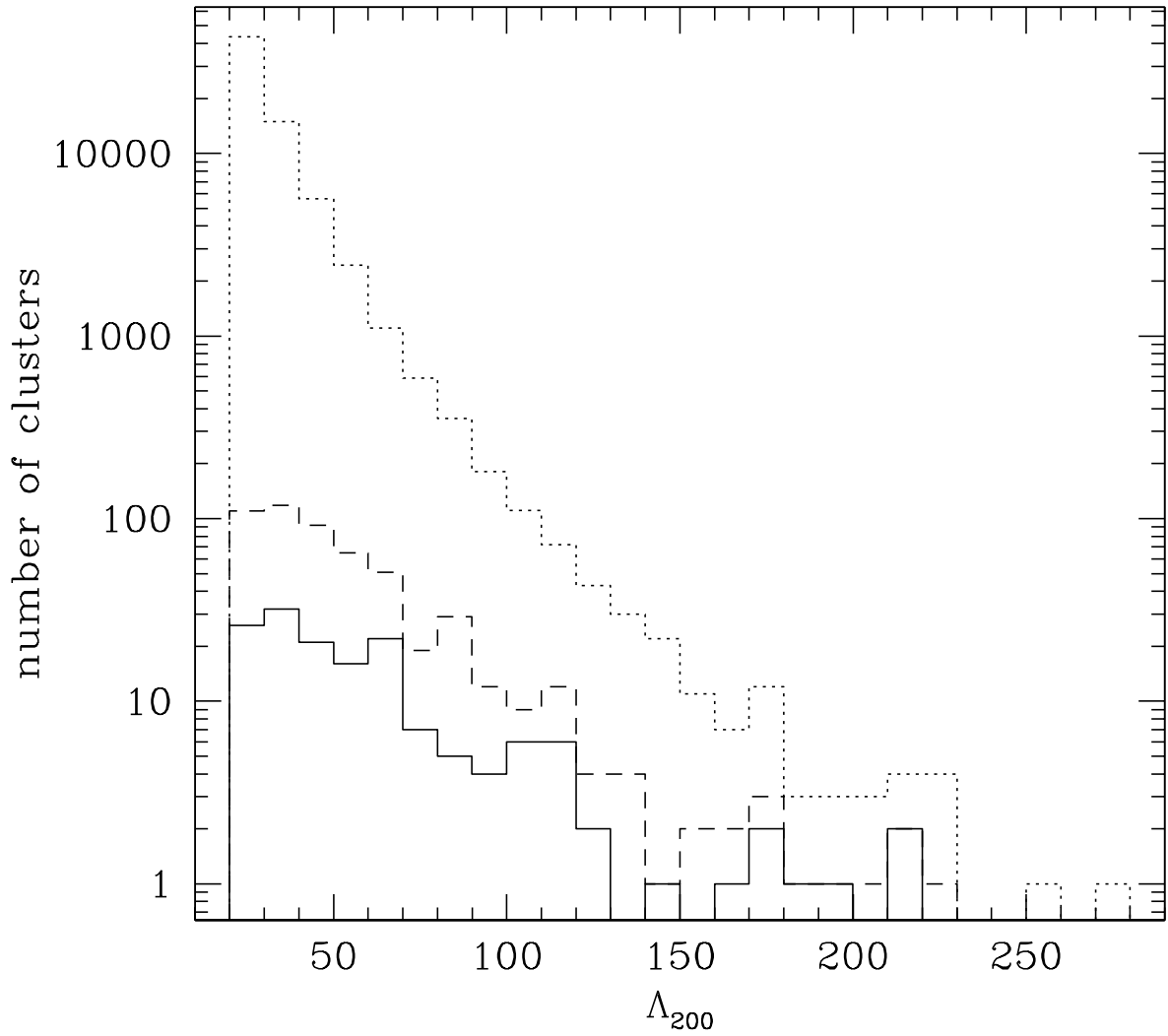}
\caption{Number of matching clusters as a function of $\Lambda_{200}$;
dotted line: all AMF DR6, dashed: BAX matches, solid: NORAS matches.
\label{xlmatch}}
\end{figure}

\begin{figure}
\plotone{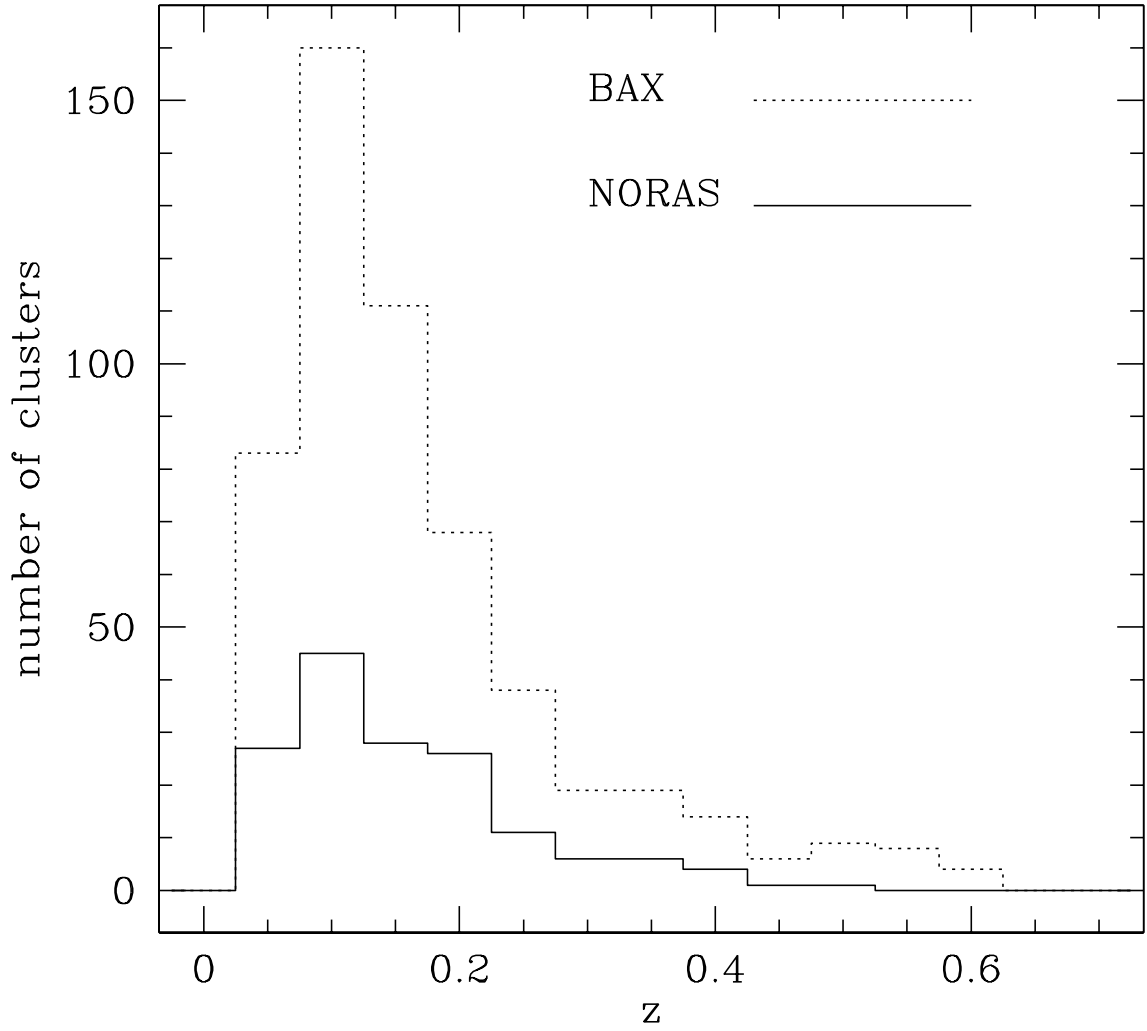}
\caption{ 
Redshift distribution of the matching clusters 
for BAX (dotted) and NORAS (solid)
\label{xzmatch}}
\end{figure}

The cross-matching procedure is performed in spatial position and redshift.
In particular, we match  the BCG position in our catalog 
to the quoted X-ray center,
which is 
typically separated by less than 100 kpc  \citep[e.g.]{linmohr04, cavagnolo09}. 
We match 539 clusters from the input list, with the  
vast majority
showing agreement between the BCG position
and the X-ray center within ~0.5 h$^{-1}$ Mpc
(Fig. \ref{bax_amf_sep}). 
Among those, 505 are  nearer than $z=0.4$. 
As for NORAS, we match 155 clusters among which 150 have a redshift 
below $z=0.4$.
The richness and redshift distribution of the matching 
clusters are reported in Figs.~\ref{xlmatch}~and~\ref{xzmatch}.


Other attempts to cross-match optical cluster catalogs with X-ray
selected ones yielded similar results. For instance, \citet{koester07}
\citep[see also][]{rykoff08} limited 
their analysis to 99 clusters from NORAS and report 
that the fraction of clusters whose
X-ray center disagrees with the optical center 
(chosen to be the BCG position) by more than 100 h$^{-1}$ kpc is
around 36\%. Similar numbers and a distribution of the positional
offset between optical and X-ray cluster centers that 
track ours can be found in \citet{lopes06}. 

\begin{figure}
\plotone{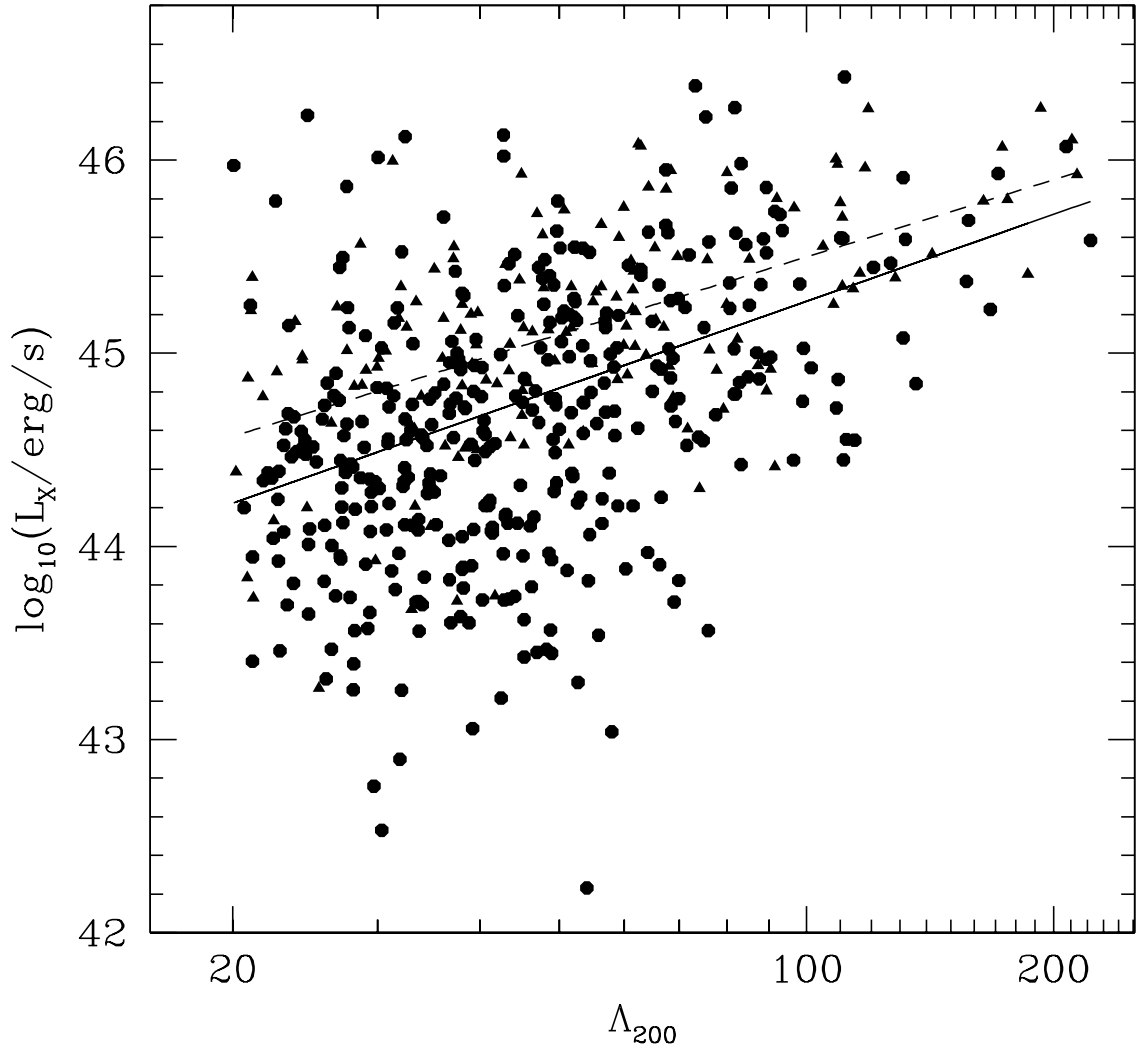}
\caption{Luminosity of BAX clusters vs. richness of 
matched AMF clusters. $L_X$ is measured in
the 0.1-2.4 keV band. Triangles: NORAS clusters, 
circles: BAX.  Solid line: fit to all data, dashed:
NORAS.
\label{bax_amf_lx} }
\end{figure}

\begin{figure}
\plotone{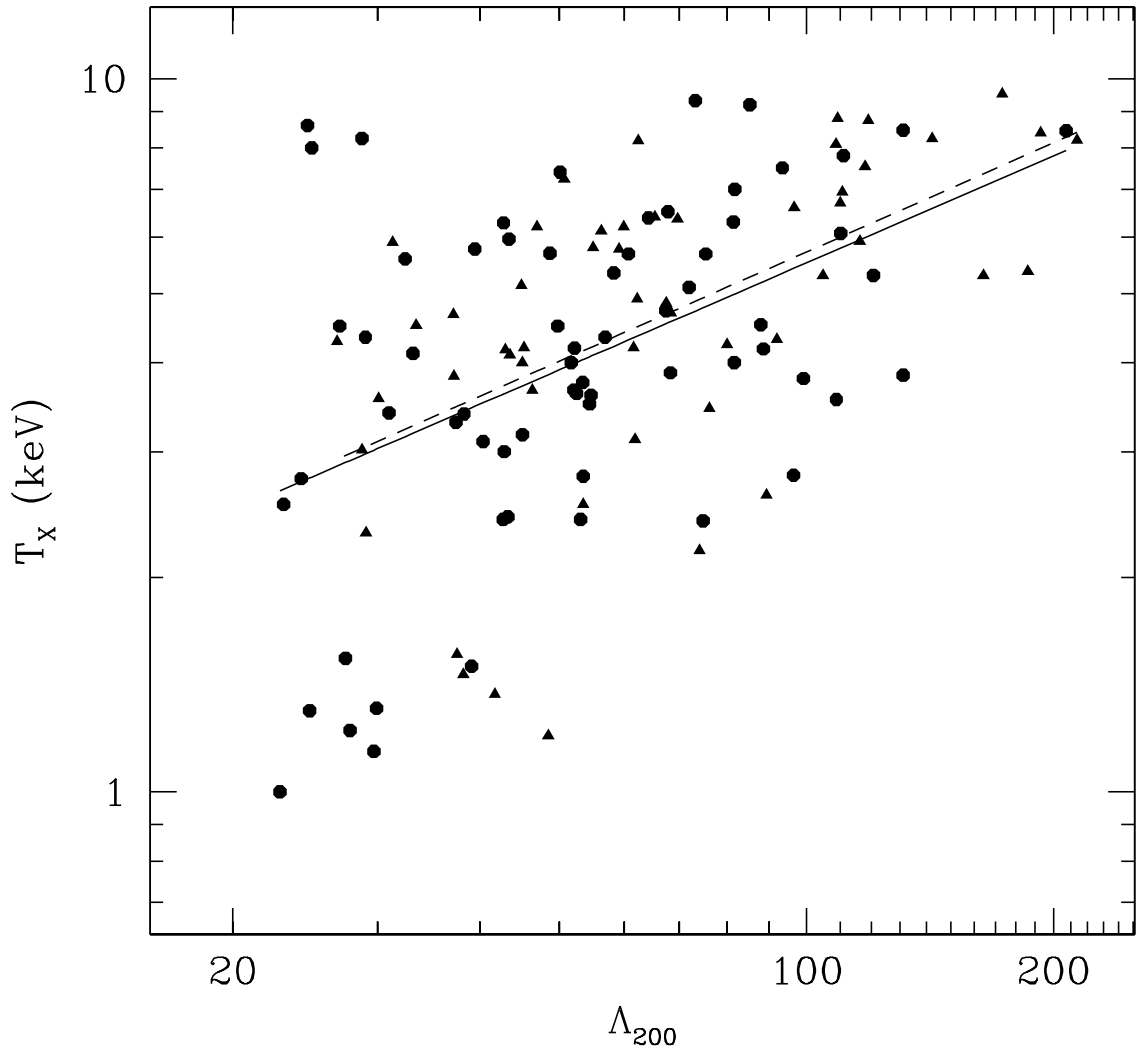}
\caption{temperature of BAX clusters vs. richness of 
matched AMF clusters.  Triangles: NORAS clusters,
circles: BAX.  Solid line: fit to all data, 
dashed: NORAS. 
\label{bax_amf_tx}. }
\end{figure}

The optical properties of matching clusters correlate 
well with the observed X--ray properties. 
In figs.~\ref{bax_amf_lx} and \ref{bax_amf_tx} 
we show X ray luminosity  and temperature versus 
richness for matching clusters.
There is a clear trend in both luminosity and 
temperature. This is somewhat different from what found 
by \citet{koester07}, who however only inspected 
 99 matches for NORAS/REFLEX on a smaller area  
and a smaller redshift range.
\citet{WHL09} look at X--ray correlations of their 
sample and find matching for about half the 
number of clusters matched with AMF.

\begin{figure}
\plotone{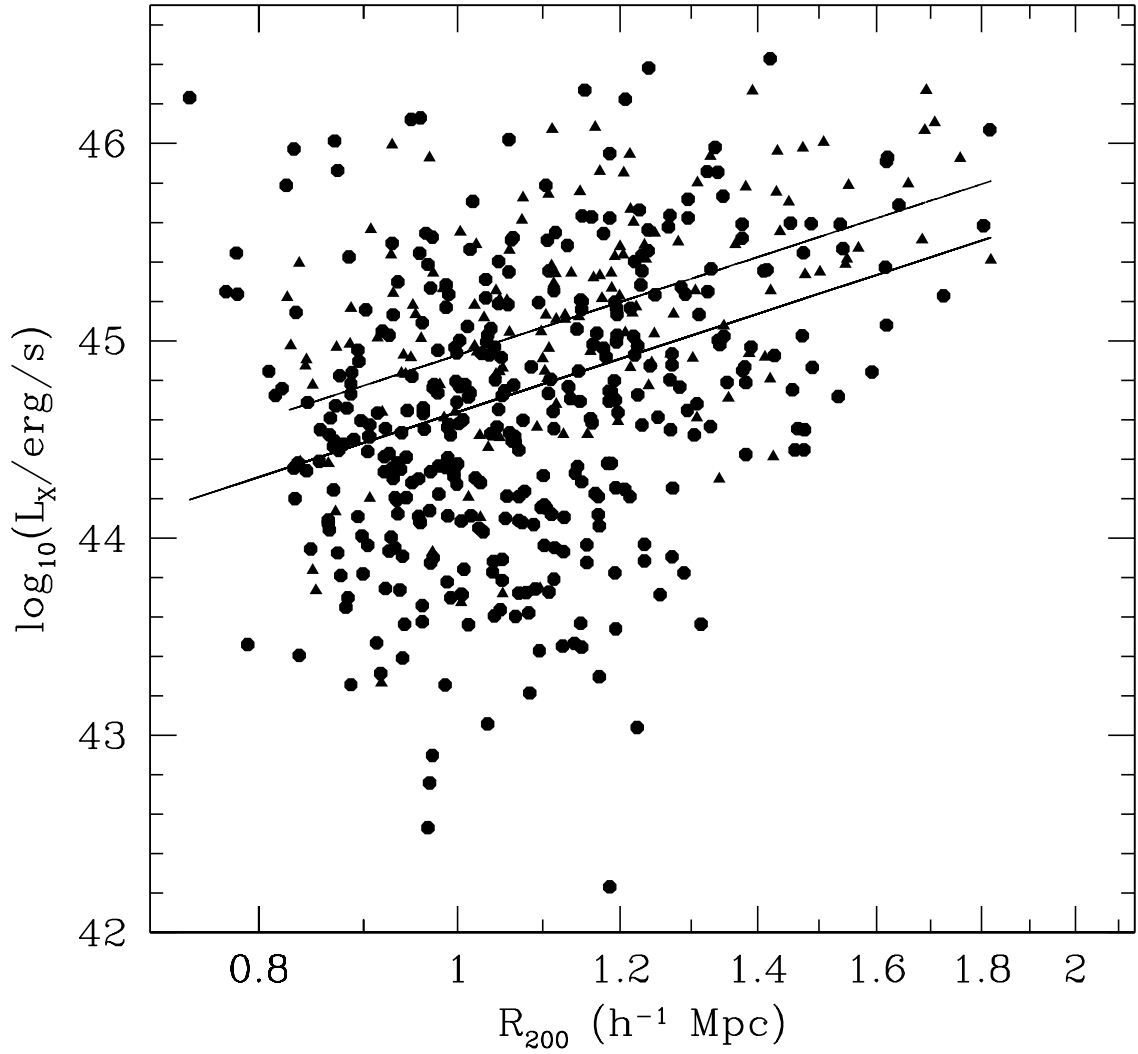}
\caption{Luminosity of BAX clusters versus radius.  
Triangles:  NORAS clusters, circles: BAX.  Solid line: 
fit to all data, dashed: NORAS. 
\label{bax_amf_lr}}
\end{figure}

\begin{figure}
\plotone{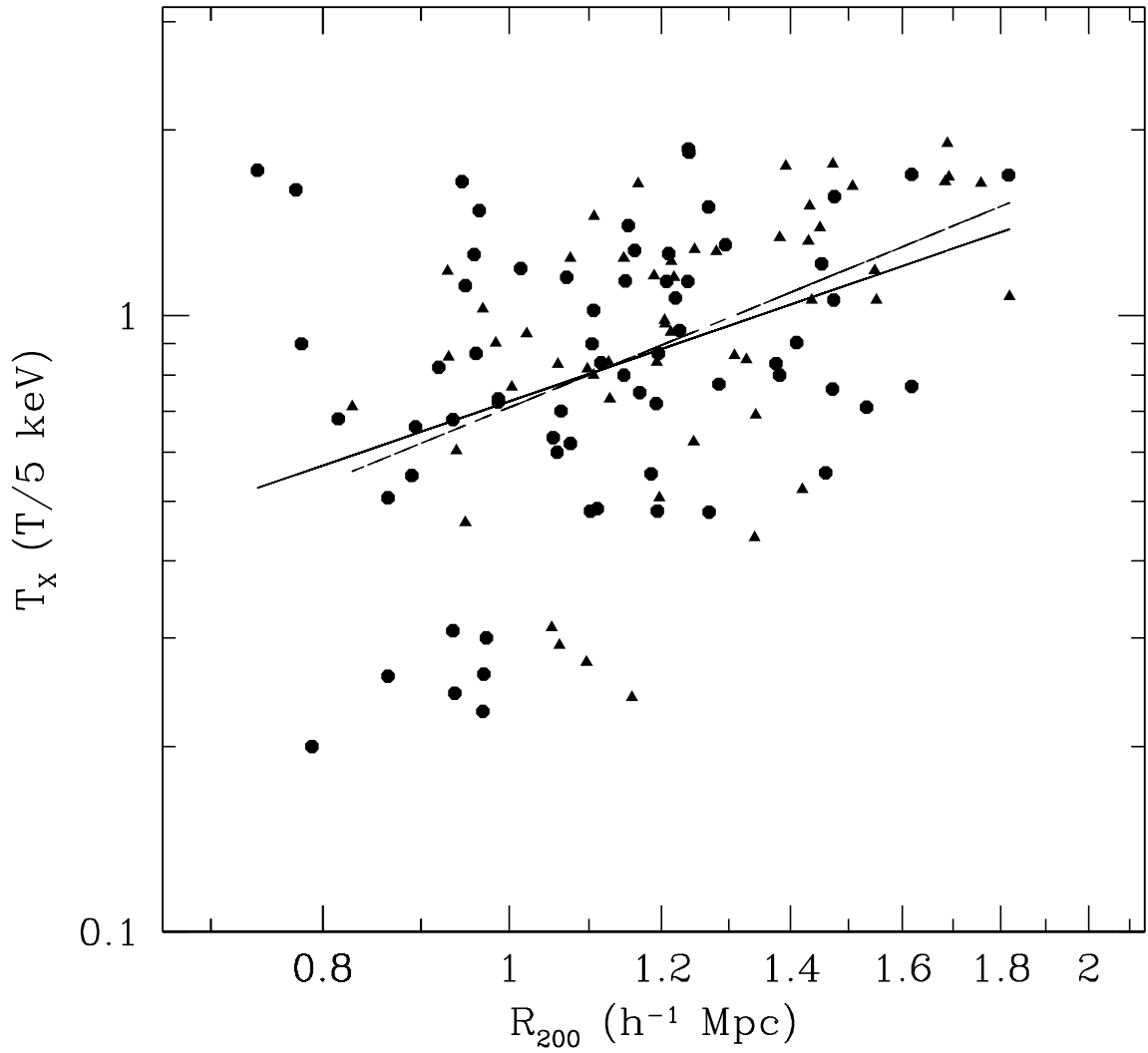}
\caption{Temperature (in keV) of BAX clusters versus radius. 
Triangles:  NORAS clusters, circles: BAX. 
Solid line: fit to all data, dashed: NORAS. 
\label{bax_amf_tr}}
\end{figure}

The scalings shown in figs.~\ref{bax_amf_lr}  and \ref{bax_amf_tx} 
are well fitted with the functions and coefficients listed in
Table~\ref{xray_coeff}.
The scatter in both relations is quite substantial 
both in temperature and  luminosity.  
(See Figures~\ref{bax_amf_lx}--~\ref{bax_amf_tr}.)
Correlation coefficients do not show significant  
difference between all BAX and flux--limited samples.

As richness correlates very well with $R_{200}$ estimates, 
the X--ray cluster properties also show 
good correlation with cluster radius.
Table \ref{xray_coeff} gives the coefficients for
the best fit lines for the X--ray measurement
vs. optical parameter graphs.
In the AMF catalog we also 
provide X--ray matching information. 

\begin{deluxetable}{lcccc}
\tabletypesize{\footnotesize}
\tablewidth{275.0pt}
\tablecolumns{5}
\tablecaption{X--ray Comparison Fitting Coefficients \label{xray_coeff}}
\tablehead{
\multicolumn{1}{c}{} & \multicolumn{2}{c}{All BAX matches} &
\multicolumn{2}{c}{NORAS matches}\\
\multicolumn{1}{c}{} & \multicolumn{1}{c}{L$_X$} &
\multicolumn{1}{c}{T$_X$} & \multicolumn{1}{c}{L$_X$} &
\multicolumn{1}{c}{T$_X$} \\
\multicolumn{1}{c}{} & \multicolumn{1}{c}{erg s$^{-1}$} &
\multicolumn{1}{c}{keV} & \multicolumn{1}{c}{erg s$^{-1}$} &
\multicolumn{1}{c}{keV}
}
\startdata
$\Lambda_{200}$, A & 42.28 & -0.25 & 42.84 & -0.26 \\
$\Lambda_{200}$, B & 1.49 & 0.498 & 1.33 & 0.508 \\
R$_{200}$, A & 44.64 & -0.14 & 44.93 & -0.15 \\ 
R$_{200}$, B & 3.40 & 1.08 & 3.40 & 1.28 \\ 
\enddata
\tablecomments{Fits to the data are of the form
$log_{10}(X) = A + B log_{10}(Y)$, where X is the
X--ray observable and Y is the optical observable.
R$_{200}$ values are always in terms of 1 h$^{-1}$ Mpc.
For fits of T$_{X}$ vs. R$_{200}$, the X--ray value
that is fit is T$_{X}$/5~keV.
}
\end{deluxetable}

\section{Conclusions \label{conc_sect}}

We present a new optical catalog of 69,173 galaxy clusters
extracted from the SDSS DR6. The catalog extends from 
z=0.045 to 0.78 on  an area of 8,420 deg$^2$.
The catalog was constructed using a maximum likelihood 
technique based on a matched filter approach which   
allows for the simultaneous determination of richness,  
core radius, redshift, with associated errors. 
The technique does not  rely heavily  on the presence of a 
luminous central red galaxy in order to detect a cluster, 
potentially allowing for the detection of clusters that 
do not present such a feature.
In this paper, we also present the catalog of the 
three brightest galaxies associated with the clusters.

We find increasing number of clusters out to 
$z \simeq 0.45$, with a slope compatible with 
the expected number  for a standard cosmology with 
$\sigma_8 \simeq 0.9$.
Richness estimates correlate well with the 
radius of the clusters ($R_{200}$).

We provide a comparison with the existing maxBCG 
catalog on DR5  and $z \le 0.3$ showing  that the two 
catalogs  only overlap at the 50\% level over the 
whole range of redshifts.
Matching between the two catalogs is high (80\%--100\%) 
for rich systems ($N_{gal}$=100 or $\Lambda_{200} > 100$) 
but is highly suppressed for smaller ones. Moreover, 
even where a match is 
found, the relation between the two richness estimates 
is not very tight and the cluster centers determined by the 
methods can be far apart.
Multiple reasons can generate these discrepancies, 
including, but not limited to, the definition of what a BCG is.

We compare our catalog with the WHL catalog which
was also built with SDSS DR6 data.  We find one-to-one
matches for $>70\%$ of WHL clusters with a richness,
$R > 25$, however clusters from the AMF catalog with
$\Lambda_{200} >$~120 match uniquely with $<~60\%$ of WHL
clusters due to the tendency of the WHL finder method
to fragment some richer AMF clusters.  WHL clusters
have one-to-one matches with AMF clusters at a rate
of 55$\%$ for all redshifts, while 30$\%$ of AMF
clusters have a unique WHL match when averaged
over redshift.

When we compare the AMF and GMBCG catalogs in the overlapping
regions, we find one-to-one matches for $>60\%$ of AMF
clusters with $\Lambda_{200} >$~80 and GMBCG clusters
with $N_{gals} >$~55.  The catalogs agree on the redshift
of approximately 5,000 clusters to a margin of $|\Delta z | <$~0.005.
The percentage of unique matches from each catalog is
in the 30$\%$ to 40$\%$ for the overlapping portion of
the redshift of the two catalogs (0.10 $\le  z \le $ 0.55).
As is the case with both the maxBCG catalog and the WHL
catalog, the GMBCG finder tends to split clusters
that the AMF finder considers as one cluster.

We cross--match the new optical catalog with X--ray  
detected cluster samples, finding 539 matches, 155 of which 
with flux--limited X--ray samples (NORAS).
We find good correlation between optical richness 
and both X--ray luminosity and temperature, 
with the same correlation found for flux limited and non flux-limited samples. 

The present catalog is sufficiently broad and deep that 
it can be used for statistical studies of 
structure formation and for comparison with cluster 
finders in other wavelengths. In particular, we anticipate 
it can be used to assess redshift information for 
Sunyaev--Zel'dovich clusters detected by 
the Planck satellite in the Northern sky.

Further  developments of this work will include 
the extension of the catalog to DR7 and inspection
of areas of the sky with deep SDSS observations.

\acknowledgments

TS, AP and EP  acknowledge support from NSF grant AST-0649899.
EP is also supported by NASA grant
NNX07AH59G and JPL--Planck subcontract 1290790, 
and would like to thank the Aspen Center for 
Physics for hospitality during the
final stages of this work.
This research has made use of: the SIMBAD database, operated at CDS,
Strasbourg, France;
the X-Rays Clusters Database (BAX), which is operated by the
Laboratoire d'Astrophysique de Tarbes-Toulouse (LATT),
under contract with the Centre National d'Etudes Spatiales (CNES); the
NASA/IPAC Extragalactic Database (NED) which is operated by the Jet
Propulsion Laboratory, California Institute of Technology, under
contract with the National Aeronautics and Space Administration.

Funding for the Sloan Digital Sky Survey (SDSS) has been provided 
by the Alfred P. Sloan Foundation, the Participating Institutions, 
the National Aeronautics and Space Administration, the National 
Science Foundation, the U.S. Department of Energy, the 
Japanese Monbukagakusho, and the Max Planck Society. 
The SDSS Web site is http://www.sdss.org/.

The SDSS is managed by the Astrophysical Research Consortium (ARC) 
for the Participating Institutions.

\appendix
\section{Photometric Flags for the SDSS Galaxy Sample 
\label{phot_flags_app}}
The following flags were set in our CAS Jobs queries
to the SDSS database in order to ensure that our 
data sample had good photometry.  The flags are given
entirely in {\tt TrueType} font, while the explanation
for the flags being set are preceded by a bullet ($\bullet$)
and follow each flag.

{\tt ((case when (type\_g=3) then 1 else 0 end) + 
(case when (type\_r=3) then 1 else 0 end) + 
(case when (type\_i=3) then 1 else 0 end)) > 1} \hfill \linebreak
$\bullet$ Object is of the type {\tt GALAXY}; \hfill \linebreak
{\tt and ((case when (dered\_g $<$ 11) then 1 else 0 end) + 
(case when (dered\_r $<$ 11) then 1 else 0 end) + 
(case when (dered\_i $<$ 11) then 1 else 0 end)) < 1 } \hfill \linebreak
$\bullet$ Apparent magnitude in the $r$-band, \quad corrected
for extinction, is $>$ 11; \qquad \hfill \linebreak
{\tt and ((flags\_r \& 0x10000000)!=0)} \hfill \linebreak
$\bullet$ Object detected in BINNED1; \hfill \linebreak
{\tt and ((flags\_r \& 0x800a0)=0)} \hfill \linebreak
$\bullet$ Eliminate objects with {\tt PEAKCENTER}, \qquad 
{\tt NOPROFILE} or {\tt NOTCHECKED} set; \qquad \hfill \linebreak
{\tt and (((flags\_r \& 0x400000000000)=0) or 
(psfmagerr\_r$<$=0.2))} \hfill \linebreak
$\bullet$ Eliminate objects where {\tt PSF\_FLU\_INTERP} 
is not set or {\tt PSF} magnitude error in the $r$-band 
is too small; \hfill \linebreak
{\tt and ((flags\_r \& 0x58)!=0x8 )} \hfill \linebreak
$\bullet$ {\tt BLENDED}, {\tt CHILD}, and {\tt NODEBLEND} objects 
have multiple peaks detected within them, and are candidates to
be a deblending parent; \hfill \linebreak
{\tt and ((flags\_r \& 0x40000=0) or (flags\_r \& 0x40000!=0 and 
flags\_r \& 0x80000000000=0 ))} \hfill \linebreak
$\bullet$ The object cannot be saturated, or have its center
too close to a saturated pixel; \hfill \linebreak
{\tt and r$<$22}\qquad \hfill \linebreak
$\bullet$ The apparent magnitude in the $r$-band is $<$ 22.

\section{Retrieving the Catalog}

The catalog (Table \ref{app_cat}), the error estimates for 
$\Lambda_{200}$, r$_c$, and position for each cluster (Table \ref{app_err}, 
the list of the three 
brightest galaxies per cluster (Table \ref{app_bcg}),
and the list of galaxies per cluster with
$\mathcal{L}_{i}(k) \ge$ 1.0 (Table \ref{app_mem}) will be available
upon request.  Please e-mail one of the authors
if you are interested in using them.

\begin{deluxetable}{rlllllcllllll}
\tabletypesize{\scriptsize}
\rotate
\tablewidth{580.0pt}
\tablenum{B.1}
\tablecolumns{13}
\tablecaption{Cluster Catalog \label{app_cat}}
\tablehead{
\multicolumn{1}{c}{a.b.d.f} &
\multicolumn{1}{c}{RA} &
\multicolumn{1}{c}{DEC} &
\multicolumn{1}{c}{z} &
\multicolumn{1}{c}{$\Delta\ln\mathcal{L}$} &
\multicolumn{1}{c}{$\Lambda_{200}$} &
\multicolumn{1}{c}{R$_{200}$} &
\multicolumn{1}{c}{...} &
\multicolumn{1}{c}{RA} & \multicolumn{1}{c}{DEC} &
\multicolumn{1}{c}{z} &
\multicolumn{1}{c}{WHL} & 
\multicolumn{1}{c}{BAX} \\ 
\multicolumn{1}{c}{} & \multicolumn{2}{c}{(degrees)} & 
\multicolumn{1}{c}{} & \multicolumn{1}{c}{} &
\multicolumn{1}{c}{($L^{*}$)} &
\multicolumn{1}{c}{(h$^{-1}$ Mpc)} & 
\multicolumn{1}{c}{} &
\multicolumn{3}{c}{of maxBCG match} & \multicolumn{1}{c}{match} & 
\multicolumn{1}{c}{match} 
}
\startdata
1.0.0.1 & 140.1428 & 30.4833 & 0.3763 & 307.9678 & 270.1469 & 1.864 & \nodata & - & - & - & - & - \\
 2.0.0.2 & 139.4840 & 51.7226 & 0.2845 & 238.2846 & 253.6087 & 1.890 & \nodata & - & - & - & - & - \\    
3.0.0.3 & 340.8307 & -9.5867 & 0.4778 & 196.6117 & 225.8704 & 1.686 & \nodata & - & - & - & J224319.8-093530 & - \\  
4.0.0.0 & 174.0642 & 40.0617 & 0.3627 & 226.7762 & 224.6193 & 1.760 & \nodata & - & - & - & J113615.9+400432 & - \\
5.0.0.5 & 188.5358 & 15.2114 & 0.2851 & 248.5700 & 222.0067 & 1.804 & \nodata & 188.47087 & 15.194643 & 0.275450 & J123416.3+151326 & Abell 1560 \\
\enddata
\tablecomments{In the first column, a is the rank of the cluster
in richness; b=1 if the cluster is on the edge of a stripe, b=0, otherwise;
d=1 if a cluster has at least five galaxies with spectroscopic measurements
as members, d=0 otherwise; f=a if a cluster is unique to a site, f=0 if the
cluster is the richest member on a blended site, f equals a of the richest
cluster on a site if the cluster in question is of lower richness, and
f=999999 if the cluster contains no galaxies with $\mathcal{L} \ge 1$.}
\end{deluxetable}

\newpage
\begin{deluxetable}{lccccccccccccc}
\tabletypesize{\scriptsize}
\rotate
\tablewidth{550.0pt}
\tablenum{B.2}
\tablecolumns{14}
\tablecaption{Error Ranges for $\Lambda_{200}$, r$_c$, and Position of AMF
DR6 Clusters \label{app_err}}
\tablehead{
\multicolumn{1}{c}{Cluster} &
\multicolumn{4}{c}{Ranges for $\Lambda_{200}$} &
\multicolumn{3}{c}{Ranges for r$_c$ (h$^{-1}$ Mpc)} &
\multicolumn{3}{c}{Ranges for RA (degrees)} &
\multicolumn{3}{c}{Ranges for DEC (degrees)} \\
\multicolumn{1}{c}{Rank} &
\multicolumn{1}{c}{95$\%^{(-)}$} &
\multicolumn{1}{c}{68$\%^{(-)}$} &
\multicolumn{1}{c}{68$\%^{(+)}$} &
\multicolumn{1}{c}{95$\%^{(+)}$} &
\multicolumn{1}{c}{95$\%^{(-)}$} &
\multicolumn{1}{c}{...} &
\multicolumn{1}{c}{95$\%^{(+)}$} &
\multicolumn{1}{c}{95$\%^{(-)}$} &
\multicolumn{1}{c}{...} &
\multicolumn{1}{c}{95$\%^{(+)}$} &
\multicolumn{1}{c}{95$\%^{(-)}$} &
\multicolumn{1}{c}{...} &
\multicolumn{1}{c}{95$\%^{(+)}$} 
}
\startdata
1 & 215.7585 & 238.7345 & 314.8423 & 342.1262 & 0.856 & \nodata & 
1.040 & 140.1395 & \nodata & 140.1760 & 30.4627 &
\nodata & 30.4873 \\
2 & 158.0631 & 179.6325 & 259.1696 & 286.1313 & 0.092 & \nodata &
0.333 & 139.4740 & \nodata & 139.5033 & 51.7149 &
\nodata & 51.7326 \\
3 & 143.1763 & 162.3866 & 229.6224 & 254.8358 & 0.379 & \nodata &
0.619 & 340.8284 & \nodata & 340.8500 & -9.6010 &
\nodata & -9.5755 \\
4 & 155.5172 & 175.8150 & 243.8724 & 267.7522 & 0.432 & \nodata &
1.079 & 174.0449 & \nodata & 174.0700 & 40.0459 &
\nodata & 40.0702 \\
5 & 158.4287 & 179.6705 & 250.4767 & 276.4389 & 0.515 & \nodata &
0.721 & 188.5200 & \nodata & 188.5551 & 15.1964 &
\nodata & 15.2299 \\
\enddata
\tablecomments{For our error estimates, we assume the shape of the
likelihood surface to be Gaussian.  We find the extrema of the
68$\%$ and 95$\%$ confidence ranges when varying any two of
richness, core radius, and position.  These extrema are reported
for both the 68\% and 95\% confidence ranges for the four
quantities listed in the table above.}
\end{deluxetable}

\begin{deluxetable}{rlllllllllllll}
\tabletypesize{\scriptsize}
\rotate
\tablewidth{570.0pt}
\tablenum{B.3}
\tablecolumns{14}
\tablecaption{Three Brightest BCG Candidates of AMF 
DR6 Clusters \label{app_bcg}}
\tablehead{
\colhead{a.b} &
\colhead{SDSS ID} &
\colhead{RA} &
\colhead{DEC} &
\colhead{k$_{z=0.0}$} &
\colhead{k$_{z=0.3}$} &
\colhead{k$_{z=0.5}$} &
\colhead{m$^u$} &
\colhead{...} &
\colhead{$\sigma_{m^z}$} &
\colhead{z$_{photo}$} &
\colhead{z$_{spec}$} &
\colhead{$\Lambda_{200}$} &
\colhead{M$^{r}$} \\ 
}
\startdata
1.1 & 587738947204284630 & 140.1074 & 30.4941 & -0.39217 & 0.20848 &
0.35816 & 19.21763 & \nodata & 0.01958 & 0.47638 & 1.00000 & 
270.1469 & -24.39410 \\
1.2 & 588017978339819919 & 140.2203 & 30.4797 & -0.39447 & 0.18756 &
0.32479 & 21.55961 & \nodata & 0.03421 & 0.32691 & 1.00000 &  
270.1469 & -22.97133 \\
1.3 & 588017978339819858 & 140.2966 & 30.4625 & -0.39447 & 0.20583 &
0.35874 & 22.12220 & \nodata & 0.02739 & 0.33160 & 1.00000 &
270.1469 & -22.45713 \\ 
2.1 & 587729388215337136 & 139.4726 & 51.7270 & -0.37158 & 0.13973 &
0.27418 & 21.04132 & \nodata & 0.01091 & 0.22093 & 1.00000 &  
253.6087 & -23.80616 \\
2.2 & 587729388215337173 & 139.5155 & 51.7261 & -0.34441 & 0.09874 &
0.18445 & 21.25870 & \nodata & 0.01742 & 0.22468 & 1.00000 &  
253.6087 & -22.25050 \\
2.3 & 587729388215337224 & 139.5786 & 51.7295 & -0.33982 & 0.10316 &
0.08794 & 21.64812 & \nodata & 0.02285 & 0.24272 & 1.00000 &  
253.6087 & -22.16702 \\
\enddata
\tablecomments{In the description above, a is the cluster rank in richness, 
and b is the rank of the galaxy in brightness in the $r$-band for cluster 
members.  The complete table has apparent magnitudes and their errors in 
all five bands.  The absolute magnitude in the $r$-band, M$^r$, is computed
at the redshift of the cluster using the interpolated value of 
the k-correction.  If there is no z$_{photo}$, the column is set to -1.0;
for no z$_{spec}$, the column is set to 1.0.}
\end{deluxetable}

\newpage
\begin{deluxetable}{llccccccccccccc}
\tabletypesize{\scriptsize}
\rotate
\tablewidth{600.0pt}
\tablenum{B.4}
\tablecolumns{15}
\tablecaption{Galaxy Membership as Determined by 
Likelihood Difference \label{app_mem}}
\tablehead{\colhead{rank} &
\colhead{SDSS ID} &
\colhead{RA} &
\colhead{DEC} &
\colhead{k$_{z=0.0}$} &
\colhead{k$_{z=0.3}$} &
\colhead{k$_{z=0.5}$} &
\colhead{m$^{u}$} &
\colhead{...} &
\colhead{$\sigma_{m^z}$} &
\colhead{z$_{photo}$} &
\colhead{err(z$_{p}$)} &
\colhead{z$_{spec}$} &
\colhead{err(z$_{s}$)} &
\colhead{$\mathcal{L}_{i}(k)$}
}
\startdata
1 & 587738947204219607 & 139.9842 & 30.4446 & -0.32752 &
0.12719 & 0.24940 & 22.66395 & \nodata & 0.09668 & 0.33350 & 0.10170 & 
1.00000 & 1.0000000 & 1.016546 \\  
1 & 587738947204219643 & 139.9797 & 30.4800 & -0.38423 &
0.17791 & 0.29166 & 22.70960 & \nodata & 0.09613 & 0.39693 & 0.07813 & 
1.00000 & 1.0000000 & 1.081769 \\
1 & 587738947204219662 & 139.9861 & 30.4828 & -0.38392 &
0.17251 & 0.29184 & 22.68239 & \nodata & 0.09734 & 0.42199 & 0.06025 & 
1.00000 & 1.0000000 & 1.206062 \\
1 & 587738947204284458 & 140.0525 & 30.5400 & -0.39447 &
0.18599 & 0.27622 & 22.91521 & \nodata & 0.07725 & 0.34760 & 0.05497 & 
1.00000 & 1.0000000 & 1.817760 \\
1 & 587738947204284490 & 140.1159 & 30.5296 & -0.34927 &
0.14850 & 0.11052 & 23.57531 & \nodata & 0.11600 & 0.30946 & 0.07936 & 
1.00000 & 1.0000000 & 2.263241 \\
\enddata
\tablecomments{Galaxies which have no spectroscopic redshift (or 
spectroscopic redshift error) are assigned a value of 1.0 for that
column.  Galaxies with no photometric redshift estimate (or error) 
appear with a value of -1.0.  The full table has apparent 
magnitudes and their errors
for all five bands.}
\end{deluxetable}

\section{Matching Clusters in Different Catalogs \label{match_app}}
The procedure for finding one-to-one matches between the
AMF DR6 catalog and other cluster catalogs is as follows:
i) using the centers of clusters in the AMF DR6 catalog as a basis, 
create a list of clusters from the other catalog whose centers
lie within a given radius of the AMF DR6 cluster center,
and whose redshifts are no more than 0.05 apart;
ii) separate this list into two lists -- one which takes
all AMF clusters in the first list and pairs them with 
the cluster from the other catalog which is the smallest
projected distance from its center,
and one which takes every cluster listed from the other
catalog and pairs it with the AMF cluster nearest its center;
iii) if the same two clusters (AMF and other catalog) are paired
in both lists from part ii, they are deemed a one-to-one match;
iv) if the AMF cluster in the list is blended, the largest cluster
on that site is selected for matching properties with the
cluster in the other catalog.
Other than the last step, the selection of one-to-one matches is
based on position, and not on any other properties of the clusters
involved.  

As an example of a case where a match is rejected, AMF cluster 2292
matches with both Abell 1999 and Abell 2000, based on clusters
with a center within R$_{200}$ of AMF 2292.  The angular 
separation is smaller for Abell 1999 than for Abell 2000. The match between
AMF 2292 and Abell 2000 is deemed one-to-one, and the match
with Abell 1999 is rejected.

\end{document}